\documentclass[fleqn,usenatbib]{mnras}

\usepackage[british]{babel}             

\usepackage{newtxmath,newtxtext}
\DeclareSymbolFont{operators}{OT1}{ntxtlf}{m}{n}
\SetSymbolFont{operators}{bold}{OT1}{ntxtlf}{b}{n}
\usepackage[T1]{fontenc}

\usepackage{graphicx}	
\usepackage{amsmath}	
\usepackage{xspace}
\usepackage{etoolbox}
\usepackage{xcolor}

\newcommand{\gaia}{\textit{Gaia}\xspace}
\newcommand{\tess}{TESS}
\newcommand{\Ktwo}{\textit{K2}}
\newcommand{\wise}{\textit{WISE}}
\newcommand{\teff}{\ensuremath{T_\textrm{eff}}\xspace}
\newcommand{\logg}{\ensuremath{\log g}\xspace}
\newcommand{\feh}{\ensuremath{[\mathrm{Fe}/\mathrm{H}]}\xspace}
\newcommand{\alphafe}{\ensuremath{[\mathrm{\upalpha}/\mathrm{Fe}]}\xspace}
\newcommand{\life}{\ensuremath{[\mathrm{Li}/\mathrm{Fe}]}\xspace}
\newcommand{\cfe}{\ensuremath{[\mathrm{C}/\mathrm{Fe}]}\xspace}

\newcommand{\kms}{\ensuremath{\textrm{km}\,\textrm{s}^{-1}}\xspace}
\newcommand{\ali}{\ensuremath{\mathrm{A}_\mathrm{Li}}\xspace}
\newcommand{\vbroad}{\ensuremath{v_\mathrm{broad}}\xspace}

\title[Lithium-rich GALAH giants]{The GALAH survey: A census of lithium-rich giant stars}

\author[S. L. Martell et al.]{
\parbox{\textwidth}{\raggedright
Sarah~L.~Martell,$^{1,2}$\thanks{Email: \texttt{s.martell@unsw.edu.au}}
Jeffrey~D.~Simpson,$^{1}$ 
Adithya~G.~Balasubramaniam,$^{1}$ 
Sven~Buder,$^{2,3}$ 
Sanjib~Sharma,$^{2,4}$ 
Marc~Hon,$^{1}$ 
Dennis~Stello,$^{1,2,4}$ 
Yuan\nobreakdash-Sen~Ting,$^{3,5,6,7}$ 
Martin~Asplund,$^{2,3}$ 
Joss~Bland\nobreakdash-Hawthorn,$^{2,4}$ 
Gayandhi~M.~De~Silva,$^{2,8}$ 
Ken~C.~Freeman,$^{3}$ 
Michael~Hayden,$^{2,4}$ 
Janez~Kos,$^{4,9}$ 
Geraint~F.~Lewis,$^{4}$ 
Karin~Lind,$^{10,11}$ 
Daniel~B.~Zucker,$^{2,12}$ 
Toma\v{z}~Zwitter,$^{9}$ 
Simon~W.~Campbell,$^{13}$ 
Klemen~\v{C}otar,$^{9}$ 
Jonathan~Horner,$^{14}$ 
Benjamin~Montet$^{1}$ 
and Rob~Wittenmyer$^{14}$ 
}
\\
\\
$^{1}$School of Physics, UNSW, Sydney, NSW 2052, Australia\\
$^{2}$Centre of Excellence for Astrophysics in Three Dimensions (ASTRO-3D), Australia\\
$^{3}$Research School of Astronomy \& Astrophysics, Australian National University, ACT 2611, Australia\\
$^{4}$Sydney Institute for Astronomy, School of Physics, A28, The University of Sydney, NSW, 2006, Australia\\
$^{5}$Institute for Advanced Study, Princeton, NJ 08540, USA\\
$^{6}$Department of Astrophysical Sciences, Princeton University, Princeton, NJ 08544, USA\\
$^{7}$Observatories of the Carnegie Institution of Washington, 813 Santa Barbara Street, Pasadena, CA 91101, USA\\
$^{8}$Australian Astronomical Optics, Faculty of Science and Engineering, Macquarie University, Macquarie Park, NSW 2113, Australia\\
$^{9}$Faculty of Mathematics and Physics, University of Ljubljana, Jadranska 19, 1000 Ljubljana, Slovenia\\
$^{10}$Max Planck Institute for Astronomy (MPIA), Koenigstuhl 17, 69117 Heidelberg, Germany\\
$^{11}$Department of Physics and Astronomy, Uppsala University, Box 517, SE-751 20 Uppsala, Sweden\\
$^{12}$Department of Physics and Astronomy, Macquarie University, Sydney, NSW 2109, Australia\\
$^{13}$School of Physics and Astronomy, Monash University, Clayton, VIC 3800, Australia\\
$^{14}$Centre for Astrophysics, University of Southern Queensland, Toowoomba, QLD 4350, Australia\\
}

\date{Accepted XXX. Received YYY; in original form \today}

\pubyear{2021}

\begin{document}
\label{firstpage}
\pagerange{\pageref{firstpage}--\pageref{lastpage}}
\maketitle

\begin{abstract}
We investigate the properties of 1262 red giant stars with high photospheric abundances of lithium observed by the GALAH and \Ktwo-HERMES surveys, and discuss them in the context of proposed mechanisms for lithium enrichment and re-depletion in giant stars. We confirm that Li-rich giants are rare, making up only 1.2 per cent of our giant star sample. We use stellar parameters from the third public data release from the GALAH survey and a Bayesian isochrone analysis to divide the sample into first-ascent red giant branch and red clump stars, and confirm these classifications using asteroseismic data from \Ktwo. We find that red clump stars are 2.5 times as likely to be lithium-rich as red giant branch stars, in agreement with other recent work. The probability for a star to be lithium-rich is affected by a number of factors, though the causality in those correlations is not entirely clear. We show for the first time that primary and secondary red clump stars have distinctly different lithium enrichment patterns. The data set discussed here is large and heterogeneous in terms of evolutionary phase, metallicity, rotation rate and mass. We expect that if the various mechanisms that have been proposed for lithium enrichment in evolved stars are in fact active, they should all contribute to this sample of lithium-rich giants at some level. 

\end{abstract}

\begin{keywords}
stars:abundances -- stars:evolution
\end{keywords}



\section{Introduction}
\label{sec:info}
Lithium-rich giants are a longstanding mystery in stellar evolution \citep{Burbidge1957,Wallerstein1969,Trimble1975, Trimble1991}. The first evolved star known to have a highly enhanced lithium abundance is the carbon-rich AGB star WZ Cassiopeiae, which was noted by \citet{McKellar1940,McKellar1941}, and the first red giant branch stars with elevated lithium abundance were reported by \citet{Luck82} and \citet{ws82}. Canonical stellar evolution predicts that a star with approximately Solar mass forms with the atmospheric lithium abundance that matches its local interstellar medium, and that abundance is largely preserved throughout the star's main sequence lifetime. The structure of Solar-mass main sequence stars, with radiative cores and fairly shallow convective envelopes, means that the material in the stellar atmosphere is never exposed to a high enough temperature to destroy the lithium \citep[$2.6\times 10^6$~K;][]{Gamow1933,Salpeter1955}. Then, in the first dredge-up phase \citep{Iben1965}, which happens as a star evolves from the main sequence toward the red giant branch (RGB), the convective envelope deepens dramatically. First dredge-up transports atmospheric material through the hot stellar interior, which subjects it to proton-capture fusion. This causes a sharp reduction in the surface abundances of lithium and carbon, and reduces the $^{12}$C$/^{13}$C ratio \citep[see, e.g.,][]{GSCB00,Lind2009}. As the star evolves along the RGB, there is a further sharp drop in photospheric lithium abundance at the luminosity function bump. Then at the tip of the giant branch, the helium flash causes a rapid and dramatic reconfiguration of the star as it moves to the red clump (RC), establishing a helium-burning core and a much more compact atmosphere. There is not a clear and well-known effect on surface abundances due to the helium flash, though it is reasonable to expect light elements such as lithium to be affected if there is any transport between the surface and the hydrogen-burning shell during this transition.

First dredge-up is a universal event in low-mass stellar evolution, and so we would expect to observe low photospheric lithium abundances in all red giant stars after this stage. However, a small fraction of giant stars, roughly one per cent \citep{Gao2019}, have high photospheric lithium abundances, and some even exceed the primordial lithium abundance \citep[e.g.][]{Yan2018}. Previous studies have uncovered a complex population of lithium-rich giants across a range of evolutionary phases and throughout the Local Group \citep[references include][]{Kraft1999,Pilachowski2000,Gonzalez2009,Kirby2016}. While the first lithium-rich first-ascent red giant branch star was discovered in a globular cluster \citep{ws82}, only a small number of additional lithium-rich globular cluster stars have been discovered \citep[e.g.][]{Kirby2016}. A fraction of lithium-rich giants exhibit features such as high rotational velocity \citep[e.g.,][]{Charbonnel2000}, or infrared excess in their spectral energy distributions \citep[e.g.,][]{Rebull2015}, but as a rule they have not been observed to differ in any systematically significant way from lithium-normal giants with the same stellar parameters and evolutionary phase \cite[e.g.,][]{MS13,Casey2016,Smiljanic2018,Deepak2019}.

Red giant branch stars that are not lithium-rich show a steady depletion in lithium abundance as they ascend the giant branch \citep{Lind2009}. This is generally interpreted as a result of internal mixing and can be modelled as a combination of rotation and thermohaline instability \citep[e.g.,][]{Charbonnel2020}. We implicitly assume that this type of depletion through internal mixing also occurs steadily in lithium-rich giants of all kinds, creating a natural timescale for re-depletion of photospheric lithium.

In response to the observational data, a number of mechanisms have been proposed for the acquisition or production of lithium in evolved stars, though there have not been many discussions of how the re-depletion rate might depend on stellar evolutionary phase or metallicity. The lithium enhancement models tend to focus on planet engulfment \citep[e.g.,][]{Carlberg2012,Aguilera-Gomez2016} or internal mixing in conjunction with the \citet{Cameron1971} lithium production process \citep[e.g.,][]{Charbonnel2000,Sackmann1992}. The models are often closely tied to particular events in stellar evolution.

More recent work \citep[e.g.,][]{Casey2019,Singh2019a} has identified that lithium-rich giants are more likely to be in the red clump (i.e., stars that are core-helium burning) than on the first ascent red giant branch (i.e., stars that are hydrogen-shell burning). The study of \citet{Kumar2020}, which is also based on GALAH data, posits a universal lithium production event at the helium flash based on the lack of red clump stars that are as lithium-poor as the stars at the tip of the red giant branch, and \citet{Yan2021} find indications of accompanying nitrogen enrichment in their large study of LAMOST data. To evaluate arguments about the source of lithium enrichment it is critical to know the evolutionary phase of the stars in question. It can be difficult to confidently separate red clump stars from red giant branch stars with similar surface gravity based on photometry or spectroscopy. Asteroseismology has the potential to provide crucial perspective on this problem, as asteroseismic parameters are clearly distinct for RGB stars with degenerate hydrogen cores and RC stars with helium burning cores \citep{Bedding2011}.

With the availability of lithium abundances from large spectroscopic projects like the \gaia-ESO Survey \citep{Gilmore2012}, the LAMOST survey \citep{Cui2012}, and the GALAH Survey \citep{deSilva15}, and the tremendous expansion in asteroseismic sky coverage from the \textit{Kepler} \citep{Stello2013}, \Ktwo\ \citep{Stello2017} and \tess\ \citep{SilvaAguirre2020} missions, we can now assemble and use large catalogues of lithium-rich giants with reliably determined evolutionary states. Recent works \citep{Singh2019a,Gao2019,Casey2019,Deepak2019,Deepak2020} have identified thousands of lithium-rich giants in the Milky Way, a major expansion from the previous small samples.

The goal of this study is to expand the parameter space of the study of lithium-rich giants. Combining the GALAH and \Ktwo-HERMES catalogues provides a large initial set of red giant stars (described in Section \ref{sec:data}). From this data set we identify red giant branch and red clump stars using a Bayesian isochrone classification scheme (Section~\ref{sec:phase}). We investigate the bulk properties of lithium-rich giant stars, including the distribution in evolutionary phase and the occurrence rate as a function of metallicity (Section~\ref{sec:teff_logg_fe_h}) and other elemental abundances (Section \ref{sec:abundances}), and we consider observational factors discussed in previous studies including rotational velocity (Section~\ref{sec:rotation}), binarity (Section \ref{sec:binarity}), and infrared excess (Section \ref{sec:ir_excess}). We close with a discussion of our findings and make the case for careful theoretical studies of the proposed pathways for lithium enrichment in giant stars (Section \ref{sec:discussion}). Appendix~\ref{app:afe} and \ref{app:orbits} consider the \alphafe distribution and kinematic properties of the stars to investigate how lithium-rich giants are distributed across Galactic populations.

\section{The data set}
\label{sec:data}

In this section we describe the overall data set (Section \ref{sec:observations}), our giant star selection (Section \ref{sec:giant_selection}), lithium abundance determination (Section \ref{sec:li_abundances}), and classification of stellar evolutionary phase (Section \ref{sec:phase}).

\subsection{Observation, reduction, and analysis}\label{sec:observations}

Our data set contains 588571 stars, of which the vast majority come from the merger of the results of the GALAH survey \citep{Martell17,Buder18} with the \Ktwo-HERMES survey \citep{Wittenmyer2018,Sharma2019}. These two projects use the same instrumental setup --- the HERMES spectrograph \citep{Sheinis15} with the 2dF fibre positioning system \citep{Lewis2002} at the 3.9-metre Anglo-Australian Telescope --- to take high-resolution ($R \sim 28000$) spectra for stars in the Milky Way. HERMES records $\sim1000$~\AA\ across four non-contiguous sections of the optical spectrum, including the region around the H$\upalpha$ line, which contains the lithium resonance line at 6708~\AA.

Each input survey has its own selection function. The main GALAH survey 75 per cent of the combined data set) uses a simple selection function to acquire a sample from which the underlying properties of the Milky Way can be straightforwardly interpreted: the target catalogue consists of all stars with $12.0<V<14.0$, $\delta < 10$~deg and $|b|>10$~deg in regions of the sky that have at least 400 targets in $\pi$ square degrees (the 2dF field of view), and all stars in the same sky region with $9.0<V<12.0$ and at least 200 stars per 2dF field of view. The \Ktwo-HERMES survey (17 per cent of the data set) is aimed at targets from the NASA \Ktwo\ mission. This survey typically observes stars in the range $10<V<13$ or $13<V<14$ with $J-K_S>0.5$. Most \Ktwo-HERMES targets observed by \Ktwo\ are from the \Ktwo\ Galactic Archaeology Program \citep{Stello2017}, which targets stars with $(J-Ks)>0.5$ \citep{Sharma2019}. A further four per cent of the data set consists of open and globular cluster targets, and the remaining four per cent is from other targets observed with HERMES that were not part of any of these surveys.

The HERMES data were all reduced with the same custom \textsc{iraf} pipeline, which is described in \citet{Kos2017}, and analysed with the Spectroscopy Made Easy (\textsc{sme}) software \citep{Valenti1996,Piskunov2017}. The analysis is described in detail in \citet{Buder2020}, but briefly, \textsc{sme} is used to perform spectrum synthesis for 1D stellar atmosphere models. We use \texttt{MARCS} theoretical 1D hydrostatic models \citep{Gustafsson2008}, with spherically symmetric stellar atmosphere models for $\logg\leq 3.5$ and plane parallel models otherwise. \textsc{sme} calculates radiative transfer under the assumption of local thermodynamic equilibrium, and so we incorporate non-LTE line formation for elements \citep[including lithium,][]{Lind2009} where the effect on abundance determination is known to be significant. In all cases the non-LTE computations are performed using the same grid of \texttt{MARCS} model atmospheres as the LTE computations.

\subsection{Giant star selection}\label{sec:giant_selection}
\begin{figure}
	\includegraphics[width=\columnwidth]{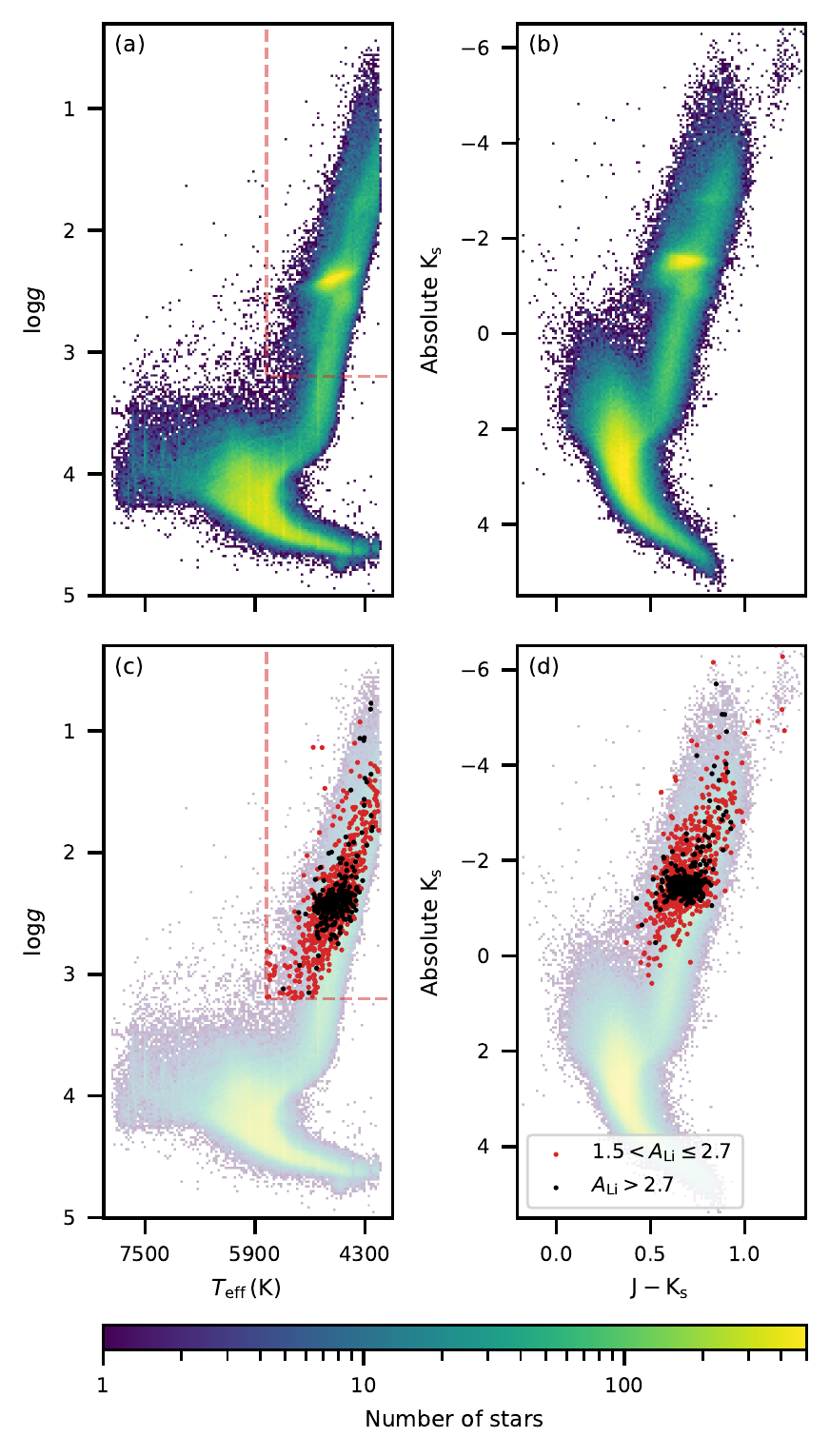}
    \caption{Kiel diagrams (left column) and absolute colour-magnitude diagrams (right column) for the ``good'' sample (i.e., no flagged problems) of 
    stars considered in this work. The dashed red rectangle in (a) and (c) shows the \teff, \logg selection used to identify giant stars. In the bottom row (c, d) we highlight giant stars with $\ali>\{1.5,2.7\}$, using red and black points respectively. This shows that Li-rich giants are found at all parts of the giant branch, but the very Li-enhanced stars ($\ali>2.7$) tend to be found in the red clump region.}
    \label{fig:teff_logg}
\end{figure}

For this work, we are using the GALAH Data Release 3 catalogue of stellar parameters and abundances. This contains 588571 stars, all of which are found in \gaia DR2 \citep{GaiaCollaboration2016,GaiaCollaboration2018b} and most of which are in the \textit{AllWISE} catalogue \citep[96 per cent;][]{Wright2010,Mainzer2011}. This cross-match used the \texttt{gaiadr2.allwise\_best\_neighbour} table created by the \gaia mission team.

We apply a number of selections in data quality and stellar parameters to identify a sample of reliable lithium-rich giant stars. We require that each star has:
\begin{itemize}
    \item the GALAH flag $\texttt{flag\_sp}==0$: no problems noted in the input data, reduction, or analysis;
    \item the GALAH flag $\texttt{flag\_fe\_h}==0$: no problems noted in the iron abundance determination;
    \item a calculated $E(B-V) < 0.33$, to avoid known difficulties with the analysis of highly extinguished spectra \citep[as discussed in][]{Buder2020};  
    \item a photometric measurement in the \wise\ $W_2$ band with a data quality flag of A, B, or C, as the WISE photometry is used as part of the identification of RC stars.
\end{itemize}
We also excluded stars in the SMC or LMC based on their spatial and kinematic properties. The LMC selection includes all stars within 5~deg of $(\mathrm{RA},\mathrm{Dec})=(78.00,-68.08)$~deg, with a proper motion within 1.5~mas\,yr$^{-1}$ of $\mu_\mathrm{RA},\mu_\mathrm{Dec}=(1.80,0.25)$~mas\,yr$^{-1}$ and a radial velocity larger than 215~\kms. The SMC selection includes all stars within 5~deg of $(\mathrm{RA},\mathrm{Dec})=(11.83,-74.11)$~deg, with a proper motion within 1.5~mas\,yr$^{-1}$ of $\mu_\mathrm{RA},\mu_\mathrm{Dec}=(0.85,-1.20)$~mas\,yr$^{-1}$, a radial velocity larger than 80~\kms, and a parallax $\varpi<0.08$~mas. These criteria retained 66.3 per cent (390004/588571) of the sample as ``good'' stars. Kiel and colour-magnitude diagrams of these stars are shown in Figure~\ref{fig:teff_logg}.

We restrict ourselves to giant stars, selecting those stars found to have effective temperature in the range $3000\mathrm{~K}\leq \teff \leq 5730\mathrm{~K}$ and surface gravity in the range $3.2 \geq \logg \geq -1.0$ (red rectangle on Figure~\ref{fig:teff_logg}a,c). Of our sample of ``good'' stars, 28.0 per cent (109340/390004) were identified as giant stars.

\subsection{Lithium abundances}\label{sec:li_abundances}
\begin{table*}
\caption{We identify 1262 Li-rich giants with reliable evolutionary stage classifications. Here we give their \gaia DR2 \texttt{source\_id}, sky locations, and GALAH DR3 stellar parameters and spectroscopic information. The full version of the table is available online; the first five entries included here are the Li-rich stars shown in Figure \ref{fig:comparison_spec_plots}, and the sixth entry is the star with the highest \ali in our sample. 
}
\begin{tabular}{rrrrrrrrr}
\hline
\gaia DR2 source\_id & RA & Dec & RV (\kms) & \teff (K) & \logg & \feh & \ali & RC or RGB\\
\hline
5371899834025124608 & $175.501$ & $-48.171$ & $1.43\pm0.67$ & $4285\pm158$ & $1.75\pm0.26$ & $-0.22\pm0.17$ & $2.21\pm0.22$ & RGB\\
6100901881763791232 & $223.933$ & $-42.034$ & $-18.00\pm0.34$ & $4429\pm80$ & $2.01\pm1.20$ & $-0.18\pm0.05$ & $2.40\pm0.08$ & RGB\\
6235140814020759808 & $236.883$ & $-25.259$ & $-23.43\pm0.56$ & $4681\pm127$ & $2.31\pm0.26$ & $-0.18\pm0.11$ & $2.25\pm0.16$ & RC\\
6129493448995721984 & $180.111$ & $-50.437$ & $-50.04\pm0.33$ & $4980\pm80$ & $2.55\pm0.21$ & $-0.21\pm0.05$ & $2.25\pm0.08$ & RC\\
3155263089390175872 & $109.992$ & $9.198$ & $38.66\pm0.59$ & $5147\pm127$ & $2.72\pm0.24$ & $-0.18\pm0.11$ & $2.29\pm0.16$ & RC\\
6221353316163376768 & $217.790$ & $-28.496$ & $-115.60\pm0.41$ & $4210\pm89$ & $0.77\pm0.45$ & $-1.67\pm0.07$ & $4.80\pm0.11$ & RGB\\
\hline
\label{tab:params}
\end{tabular}
\end{table*}

\begin{figure*}
	\includegraphics[width=\textwidth]{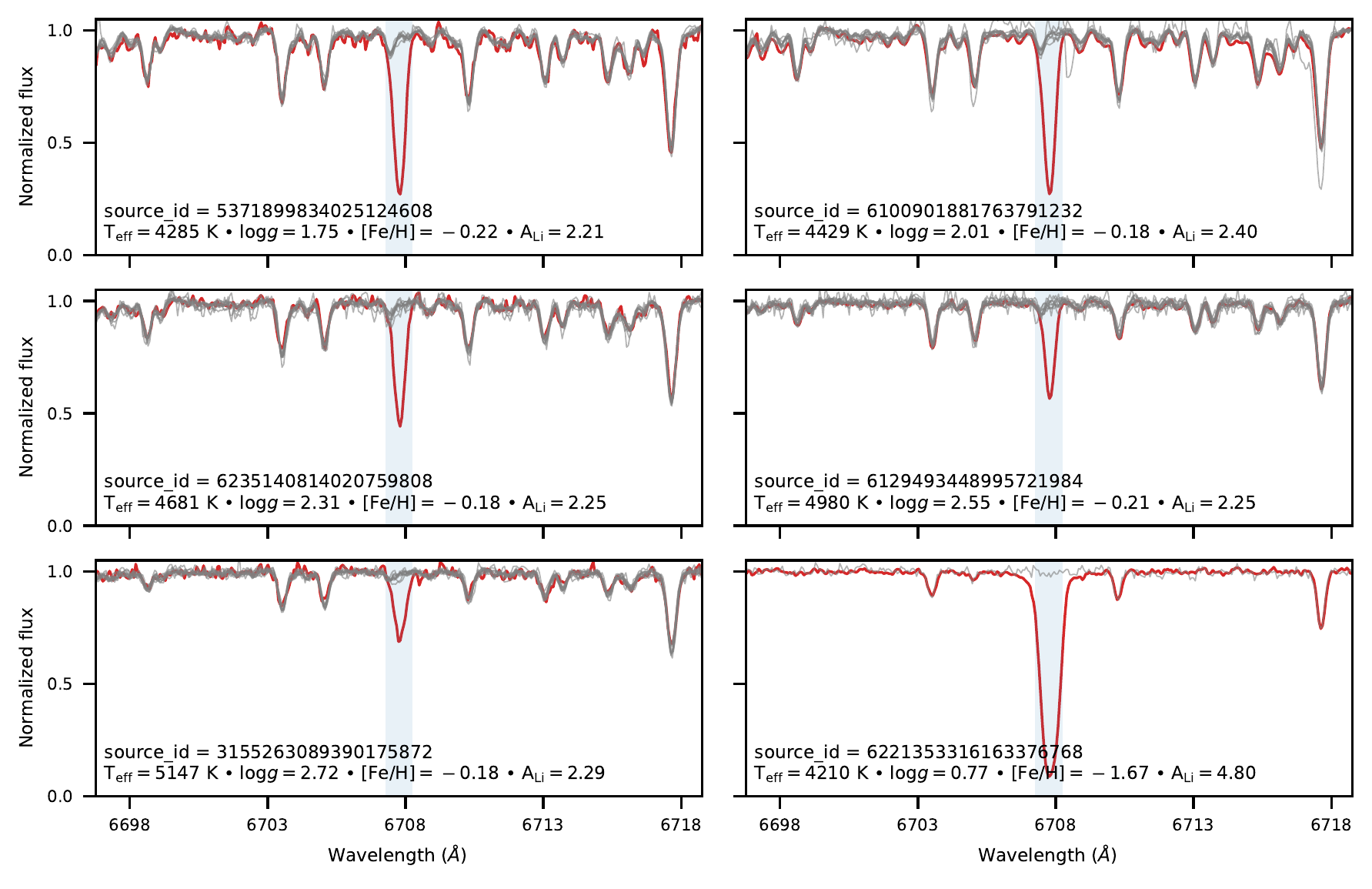}
    \caption{Examples of the spectral region containing the Li~6708~\AA\ line (indicated with the shaded blue region in each panel), as observed with HERMES, for stars from a range of \teff and \logg along the giant branch. In each panel we highlight one Li-rich giant star (red line; also the \texttt{source\_id} and stellar parameters) and up to 10 other randomly selected stars with similar stellar parameters (grey lines) --- namely $\Delta(\teff)<50$~K, $\Delta(\logg)<0.2$, $\Delta(\feh)<0.03$, $\Delta(\alphafe)<0.05$, $v_\mathrm{broad}<5$~\kms. In all panels except for the bottom-right, the stars have approximately the same metallicity and \ali. The bottom-right panel highlights the most Li-rich star in the sample. There are two things to note: first, for most of these giant stars, the lithium line is either weak or not visible; second, for a given \ali, the lithium absorption line gets weaker with increasing \teff\ --- which means that the minimum detectable \ali in our data set is higher at higher temperature.}
    \label{fig:comparison_spec_plots}
\end{figure*}

The lithium abundance of each star was determined as part of the main analysis of the GALAH and \Ktwo data sets from synthesis of the 6708~\AA\ lithium line. We report the lithium abundance value in the form of $\ali(\equiv\life+\feh+1.05)$, where the $\mathrm{A}_\mathrm{X}$ abundance scale gives the number density of element X on a logarithmic scale relative to hydrogen, with $\mathrm{A}_\mathrm{H}= 12$ by definition, and 1.05 is the lithium abundance of the Sun reported in \citet{Asplund2009}. We follow the typical convention from the literature of considering a giant star to be lithium-rich if its abundance $\ali>1.5$, and note that \citet{Kirby2016} and \citet{Charbonnel2020} discuss whether this Li-rich limit should be a function of stellar parameters. We also highlight throughout this work the subset of these Li-rich giant stars with \ali above the primordial value of 2.7 \citep{Cyburt2008,Fields2019}, because it serves as a useful landmark in the abundance space.

Each elemental abundance in the GALAH DR3 catalogue has an associated reliability flag. For this work we require  $\texttt{flag\_li\_fe}==0$, which indicates a significant line detection with no flagged problems. Of the 109340 ``good'' giant stars, 10756 stars (9.8 per cent) met this criterion. As with any spectral line, the strength of the lithium line is a complicated function of the stellar parameters and the lithium abundance of the star. In Figure \ref{fig:comparison_spec_plots} we show example HERMES spectra for giant stars of similar \feh across the range of \teff and \logg values covering the giant branch. Each panel shows the spectrum of one Li-rich giant (red line; in all cases $\ali\sim2.3$), compared to the spectra of 10 randomly selected stars with similar stellar parameters (grey lines). This highlights that for most giant stars, the 6708~\AA\ line of lithium is not detectable, and that the sensitivity to lithium abundance decreases as \teff rises. This explains why only 9.8 per cent of our giant star sample has a measured (and non-flagged) value for \ali, despite our high quality spectra.

Very strong lithium lines are also challenging for the GALAH data analysis pipeline to handle correctly, since at high abundance the curve of growth becomes flatter and consequently small changes in line strength would require larger changes in abundance. While our data set includes a number of stars with \ali over 4.0, we would advise caution when using abundances in this regime. Although we are certain that these stars do have strong lithium absorption features and therefore high abundances, we are less sure that the abundance values reported in the GALAH DR3 catalogue are accurate. These stars will require boutique analysis for accurate abundance determination.

For our giant star sample, 1262/109340 (1.2 per cent) have $\ali>1.5$. This is consistent with the 1.29 per cent value found independently in the LAMOST survey \citep{Gao2019}. Of our 1262 Li-rich giants, 323 stars lie above the primordial value of $\ali = 2.7$ --- we refer to these stars throughout as ``super Li-rich''.

The location of our Li-rich giants in the Kiel and color-magnitude diagrams are shown with red and black points on Figure~\ref{fig:teff_logg}. On the lower giant branch ($\logg\sim3$), there is a dearth of Li-rich stars on the cooler side of the giant branch. This has been previously observed \citep{Ramirez2012,Buder18}, and is caused by the deeper surface convective envelopes of cooler stars, which extend to hotter regions in the stellar interior and allow for more rapid depletion of the surface lithium abundance.

\subsection{Classifying evolutionary phase}
\label{sec:phase}

The evolutionary state of lithium-rich giants is an essential piece of knowledge for evaluating models to explain their enrichment. The GALAH DR3 data set contains two main populations of low-mass giants in the Milky Way:
\begin{itemize}
    \item Red giant branch (RGB) stars, on their first ascent of the giant branch, with an inert helium core and a hydrogen-burning shell. The RGB spans a wide range in \logg and luminosity.
    \item Red clump (RC) stars, in the stage directly after the first ascent of the giant branch, with a helium-burning core and a hydrogen-burning shell. RC stars occupy only a small range of He core mass and therefore luminosity, and they fall near RGB stars with the same \logg in the observable parameter space. 
\end{itemize}

About one-third of evolved stars in a magnitude-limited survey are expected to be RC stars \citep{Girardi2016}. As noted by previous authors \citep{Casey2019, Gao2019, Zhang2020}, Li-rich giants are more likely to be RC stars than RGB. As shown in the bottom row of Figure \ref{fig:teff_logg}, there is a clear over-density of stars corresponding to the location of the red clump, both for the Li-rich and super-Li-rich stars.

As discussed in Section~\ref{sec:info}, RC and RGB stars can be distinguished using asteroseismology. Our ability to infer the interior properties of stars has been greatly improved by the precise photometry recorded by various space missions (e.g., \textit{CoRoT}, \textit{Kepler}, \Ktwo, \tess). The gravity-mode period spacing ($\Delta\Pi$), the frequency offset ($\epsilon$), and the large frequency spacing ($\Delta\nu$) can be measured from the power spectra derived from these light curves, and these quantities take very different distributions for red clump and red giant branch stars \citep[e.g.,][]{Mosser2012,  Kallinger2012, Stello2013, Vrard2016}. This technique has been used for small samples of Li-rich giants to get unambiguous classifications \citep{Singh2019a,Casey2019}, and the asteroseismic quantities can be used to train data-driven methods for spectroscopic classification \citep[e.g.,][]{Hawkins2018a}.

Most of our giant stars do not have the necessary time-series photometry for asteroseismic determination of their evolutionary phase. The time series photometry in \Ktwo\
is not as extensive as it was for the original \textit{Kepler} mission, making measurements of the seismic properties more difficult to obtain, especially for $\Delta\Pi$. However, classification of red clump versus red giant branch stars can be done reliably from \Ktwo\ data using well-trained machine learning methods. For this work, we used classifications performed using the method described in \citet{Hon2018}, which has a 95 per cent accuracy. 1002 of the \Ktwo\ stars in our sample have reliable stellar parameters and seismic classifications from this technique. Of these, 579 stars are classified as RC and 423 are RGB, but only 8 of the 1002 stars are lithium-rich. We indicate the seismically-classified stars in figures throughout this work, but do not rely on them for any of the conclusions. 

Since only a small fraction of our Li-rich giants have seismic classifications, for the majority of our stars we used RGB/RC classifications from the Bayesian Stellar Parameters estimator (\textsc{bstep}). This is described in detail in \citet{Sharma2018}, but briefly, it provides a probabilistic estimate of intrinsic stellar parameters from observed parameters by making use of stellar isochrones. BSTEP results are available as a value-added catalogue in GALAH DR3. For results presented in this paper, we exploit the PARSEC-COLIBRI stellar isochrones \citep{Marigo2017}. The red-clump evolutionary state is labelled in the isochrones and we make use of this information to assign a probability for a star belonging to the red clump based on its observed stellar parameters (\texttt{is\_redclump\_bstep}). We supplement these classifications by taking advantage of the fact that RC stars are standard candles. The \wise\ $W_2$ absolute magnitude of the stars was calculated using the conventional relationship $M_\lambda = m_\lambda-5\log(\mathrm{r}_\mathrm{est})+5$, with the distance $\mathrm{r}_\mathrm{est}$ taken from \citet{Bailer-Jones2018}. The vast bulk of our RC stars were found in the range $W_2 = -1.63\pm0.80$, in line with expectations \citep{Karaali2019, Plevne2020}. Our RC and RGB selections are made as follows:
\begin{itemize}
    \item RC stars: \textsc{bstep} RC probability $\texttt{is\_redclump\_bstep}\geq0.5$ and absolute magnitude $|W_2 +1.63| \leq 0.80$,
    \item RGB stars: \textsc{bstep} RC probability $\texttt{is\_redclump\_bstep}<0.5$ or absolute magnitude outside the range $|W_2 +1.63|>0.80$.
\end{itemize}

\begin{table}
\caption{Comparison of the classifications of the RC and RGB stars for those stars that have classifications from both \citet{Hon2018} and our \textsc{bstep} selection, with recall and precision rates included. As an example, 579 stars were classified as RC stars using their seismic information and 506 of these 579 were classified as RC stars using \textsc{bstep}: an 87 per cent recall rate. Conversely, 614 stars were classified as RC by \textsc{bstep}, and the same 506 of these were classified as RC from seismic information: an 82 per cent precision rate.}
\begin{tabular}{llll}
 & RC (\textsc{bstep}) & RGB (\textsc{bstep}) & Total\\
\hline
RC (seismic) & 506 (82\%; 87\%) & 73 & 579\\
RGB (seismic) & 108 & 315 (81\%; 74\%) & 423\\
\hline
Total & 614 & 388 & 1002\\
\hline
\label{tab:rc_comparison_spec}
\end{tabular}
\end{table}

In Table \ref{tab:rc_comparison_spec} we compare the results from the seismic and isochrone-based \textsc{bstep} classifications. Considering, for instance, the red clump stars, the \textsc{bstep} classification recalls 87 per cent (506/579) of red clump stars identified from asteroseismology. For the RGB stars this recall rate is 74 per cent. Conversely, of the 614 stars classified as RC stars by \textsc{bstep}, 506 were classified as RC by the seismic method, giving a precision of 82 per cent. Similarly, for the RGB stars the precision is 81 per cent.

\begin{figure}
	\includegraphics[width=\columnwidth]{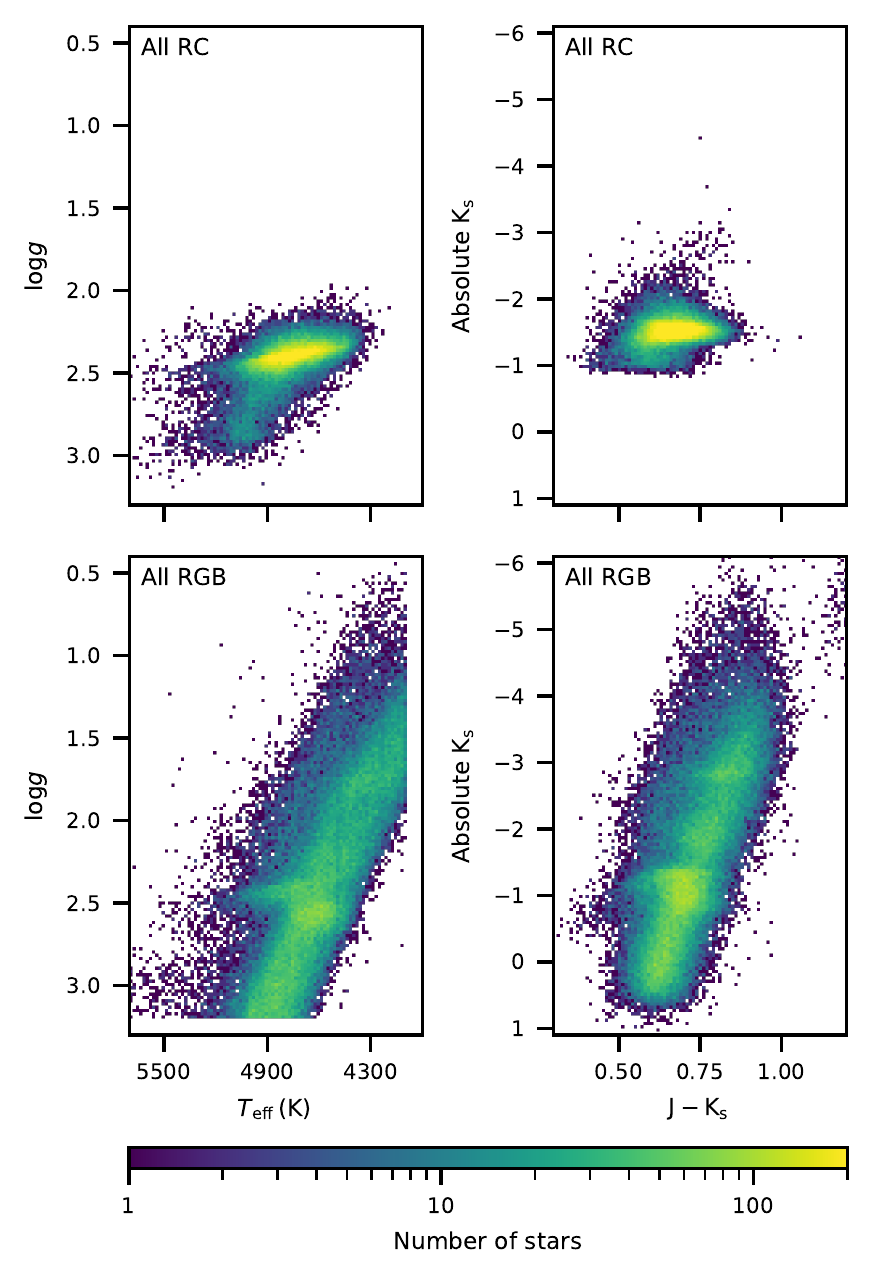}
    \caption{Kiel diagrams (left column; \teff vs \logg) and absolute colour-magnitude diagrams (right column) for the stars selected to be from (top row) RC, and (bottom row) RGB. For the RC stars both the primary RC and the higher \logg secondary RC can be seen. In the RGB panels the region of higher density on the lower giant branch is the RGB bump.}
    \label{fig:teff_logg_RC_RGB}
\end{figure}

\begin{table}
\caption{Counts of RGB and RC stars from our \textsc{bstep} classification, and the number of stars that are Li-rich ($\ali>1.5$) and super Li-rich ($\ali>2.7$). Using the mass estimates from \textsc{bstep}, stars below 1.7 
$M_{\odot}$ 
are classed as primary RC (pRC) stars, and stars above this mass as  secondary RC (sRC) stars. Where there are two percentages in parentheses, the first shows the proportion with respect to that of the total population of that type --- e.g., for Li-rich RC stars, $726/1262=58$ per cent. The second shows the percentage of the previous column value, e.g., there are 109 super Li-rich RGB stars, which is 20 per cent of the total number of Li-rich RGB stars (536).}
\begin{tabular}{lrrrr}
\hline
Star type & Total stars & Li-rich & Super Li-rich\\
\hline
All giants & 109340 & 1262 (1.2\%)& 323 (25.6\%)\\
RGB & 70343 (64\%) & 536 (42\%; 0.76\%) & 109 (34\%; 20\%)\\
RC & 38997 (36\%) & 726 (58\%; 1.9\%) & 214 (66\%; 29\%)\\
\quad \textit{pRC} & 34938 (32\%) & 580 (46\%; 1.7\%) & 195 (60\%; 34\%)\\
\quad \textit{sRC} & 4059 (3.7\%) & 146 (12\%; 3.6\%) & 19 (5.9\%; 13\%)\\
\hline
\label{tab:rc_rgb_counts}
\end{tabular}
\end{table}

Using the \textsc{bstep} classification of RC and RGB stars for our full set of 109340 giants, 38997 (36 per cent) are on the RC, and 70343 (64 per cent) belong to the RGB --- as expected for a magnitude-limited survey \citep{Girardi2016}. These results are presented in Figure \ref{fig:teff_logg_RC_RGB} and in Table~\ref{tab:rc_rgb_counts}. The expected morphologies in the Kiel and absolute colour-magnitude diagrams are recovered; namely, in the Kiel diagram, the RC can be divided into the primary red clump and the secondary red clump that consists of slightly more massive stars, and the RGB shows evidence for the luminosity function bump at a slightly higher \logg than the bulk of the RC. 

One of the most informative observables for Li-rich giants is the distribution in evolutionary phase, since different models for Li enrichment may be tied to particular processes --- for example, planet engulfment should happen during the RGB phase, while the star is expanding dramatically, but not afterward. It is important to consider this distribution not just in the sheer number of stars at a particular evolutionary phase, but rather in terms of the probability for a star at a given phase to be Li-rich. However, this statistic has not been reported previously. We find that p(\textrm{Li|RC}), the probability for an RC star to be Li-rich, is 1.9 per cent (726/38997). This is more than twice as large as p(\textrm{Li|RGB}), the probability for an RGB star to be Li-rich (536/70343; 0.8 per cent). This implies complexity in the process of Li enrichment, perhaps indicating that it occurs in a larger fraction of RC stars than RGB stars, or that there is a longer timescale for re-depletion of Li from the atmosphere in RC stars, or that the shorter lifetime of the red clump phase is more similar to the lithium depletion timescale than the lifetime of the red giant branch phase is. Assembling a thorough sample of the intrinsic and observable properties of Li-rich and Li-normal giants is crucial for constraining and improving models for Li enrichment.

\section{Lithium-rich giants as a stellar population}
\label{sec:newstuff}
In this section we compare and contrast Li-normal and Li-rich giants in the fundamental stellar parameters \teff, \logg and \feh (Section \ref{sec:teff_logg_fe_h}), elemental abundances (Section \ref{sec:abundances}), rotation rates (Section \ref{sec:rotation}), 
binarity (Section~\ref{sec:binarity}), and infrared excess (Section \ref{sec:ir_excess}). 

\subsection{Lithium-rich giants across the stellar parameter space}\label{sec:teff_logg_fe_h}
\begin{figure*}
	\includegraphics[width=\textwidth]{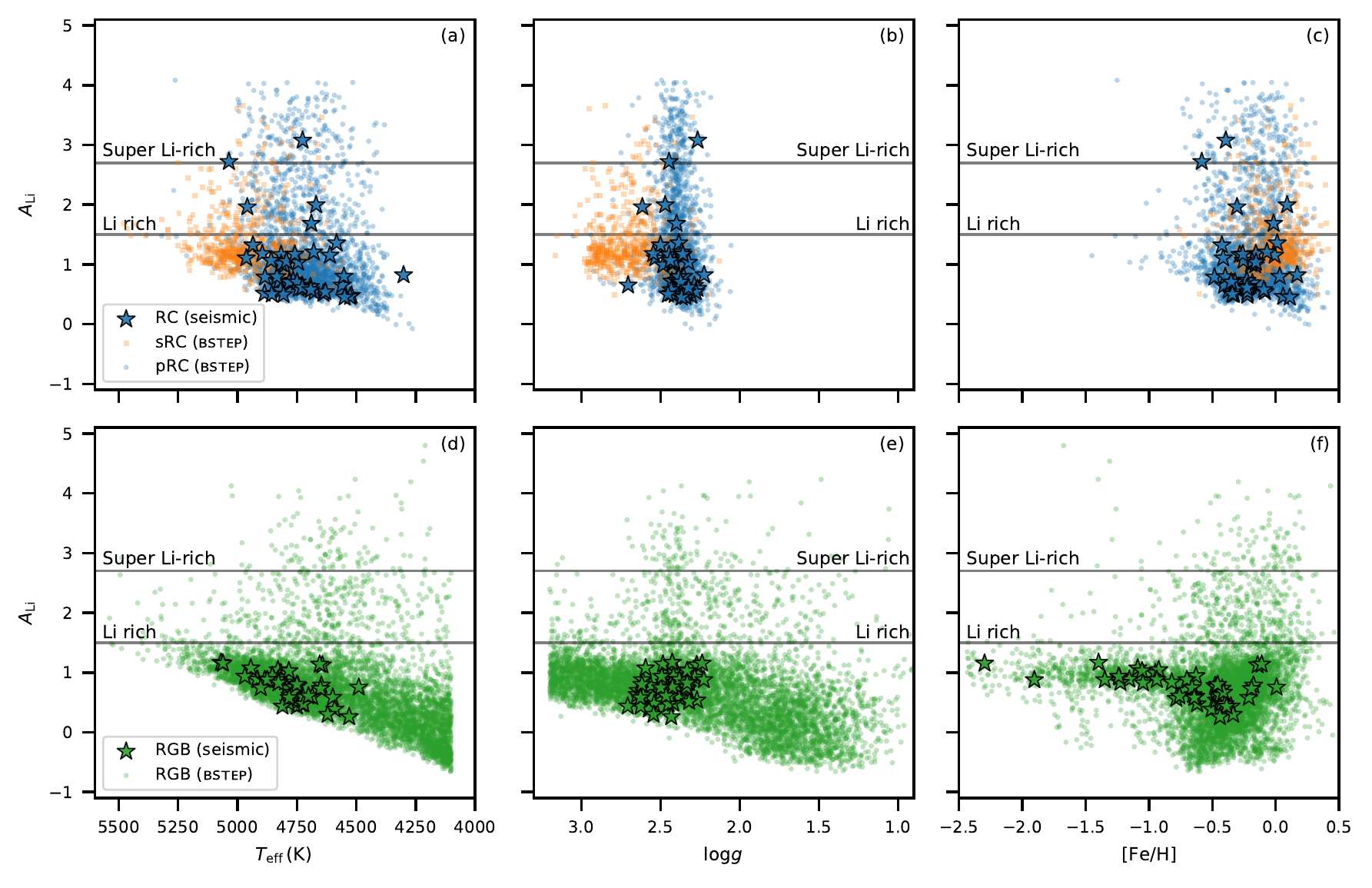}
    \caption{Comparing the \ali abundances with respect to \teff (left column), \logg (middle column), and \feh (right column). These are split into the red clump sample (upper panel of each column) and red giant branch sample (lower panel of each column). In all panels we show the stars classified as RC or RGB by the isochrone-based \textsc{bstep} method (dots) and the asteroseismic classifications (black-edged star symbols). The horizontal lines in all panels indicate criteria for whether a star is Li-rich or super Li-rich. In the \teff panels there is a clear lower envelope in \ali, indicating a detection limit driven by line strength. In the upper panels we plot stars on the primary red clump (pRC) as blue points and stars from the secondary red clump (sRC) as orange points. This highlights that the distribution in \ali is different between the two groups, with proportionally fewer super Li-rich sRC stars. In the \feh RC panel there are also very few Li-rich stars with $\feh<-1$, which is real and not a matter of sensitivity to lithium line strength.}
    \label{fig:teff_logg_fe_h_ali_rc_rgb}
\end{figure*}

Figure~\ref{fig:teff_logg_fe_h_ali_rc_rgb} presents \ali for our giants with respect to their basic stellar parameters: \teff, \logg, \feh. The upper row shows only red clump stars, and the lower row shows only red giant branch stars. The same stars are shown in all three columns, and stars with asteroseismic classifications are represented with star shapes. Horizontal lines mark the typical definition of ``lithium-rich'' at $\ali=1.5$ and the primordial lithium abundance, $\ali=2.7$.

There is a \teff-dependent lower envelope to the lithium abundance that can be measured for giant stars in our data set, which is a result of a weaker 6708~\AA\ resonance line at higher \teff, for a fixed \ali\ --- see the spectra plotted in Figure~\ref{fig:comparison_spec_plots} and the discussion in Section \ref{sec:li_abundances}. It is possible that we have failed to measure abundances for some of the hotter Li-rich giants.

Using the mass estimated for each star by \textsc{bstep}, the RC sample can be divided into pRC stars ($M_*<1.7$~ $M_{\odot}$; \citealt{Girardi2016}) and sRC stars ($M_*\geq1.7$~ $M_{\odot}$). There are 4059 sRC stars, of which 146 are Li-rich, but only 19 of these are super Li-rich --- 0.4 per cent of sRC stars and 13 per cent of Li-rich sRC stars. Meanwhile, there are 34938 pRC stars, of which 580 are Li-rich, and 195 are super Li-rich (0.6 per cent; 33.6 per cent). In the top panels of Figure \ref{fig:teff_logg_fe_h_ali_rc_rgb} these two groups are distinguished by colour, with pRC stars plotted in blue and sRC stars in orange.

The short timescales relevant to red clump stars, and the timescale differences between pRC and sRC stars, provide some useful insights into lithium enrichment and re-depletion. The helium burning lifetime of pRC stars is only $\approx 100$ Myr \citep{Girardi1999}, so if there is a universal lithium production process that occurs at the helium flash, as suggested by \citet{Kumar2020}, the lithium re-depletion must be fast enough that 98.4 per cent of pRC stars descend to $0.5 < \ali < 1.5$ in that time. At the beginning of their helium burning phase the sRC stars all have ages of $\approx 1$ Gyr because only stars in a small range of mass and metallicity pass through the sRC, and their helium burning lifetime is another 1 Gyr. A higher fraction of sRC stars exhibit lithium enrichment than pRC stars (3.6 per cent versus 1.6 per cent) despite the longer time they spend on the red clump. This suggests some difference between the lithium enrichment process in the two groups: perhaps they receive a different amount of initial lithium enrichment, or they have a different probability of experiencing lithium enrichment, or they experience a different depletion rate while on the red clump. This suggestion is strengthened by the fact that the pRC stars are more likely to be super lithium-rich than sRC stars, which acts in the opposite sense to the difference in lithium enrichment likelihood. Another possible contributor to these differences is the wider range in mass and metallicity for pRC stars compared to sRC stars, which may create or allow more diversity of behaviour within the group of pRC stars.

The luminosity function bump may play an important role in the study of Li-rich giants. As an event characterised by the introduction of fresh fuel into the hydrogen burning shell and an opportunity for increased mixing between the surface and the interior, it may be responsible for both further lithium depletion \citep[e.g.,][]{Lind2009} and lithium production through the Cameron-Fowler process \citep[e.g.,][]{Sackmann1992, Mazzitelli1999, Sackmann1999, Charbonnel2000}, depending on the exact conditions. There does appear to be a dearth of Li-rich RGB stars at the high-gravity end of the RGB star distribution. However, this picture is complicated, as stars below the RGB bump will also be hotter and therefore closer to the lithium detectability limit.

The right column of Figure \ref{fig:teff_logg_fe_h_ali_rc_rgb} shows the behaviour of \ali with metallicity for our giant stars, separated into the RC and RGB populations. The RC cohort clearly lacks stars with $\feh<-1$, which is to be expected for red clump stars in the Milky Way at the current age of the Galaxy --- there is a minimum mass for RC stars, and as a result they are a moderately young and metal-rich population \citep[e.g.,][]{Ramirez2012}. The evolution of metal-poor RC stars is also faster than for more metal-rich RC stars \citep{Girardi2016}, making them less likely to be observed. 

The metal-poor RGB stars are almost exclusively below the Li-rich threshold, with a concentration at \ali$\approx 1.0$. These stars are mainly located near the luminosity function bump, indicating that first dredge-up has reduced their \ali abundance to around 1, and it will continue to fall as they evolve along the RGB as a consequence of deep mixing \citep[e.g.,][]{Lind2009,Angelou2015}.

\begin{figure}
	\includegraphics[width=\columnwidth]{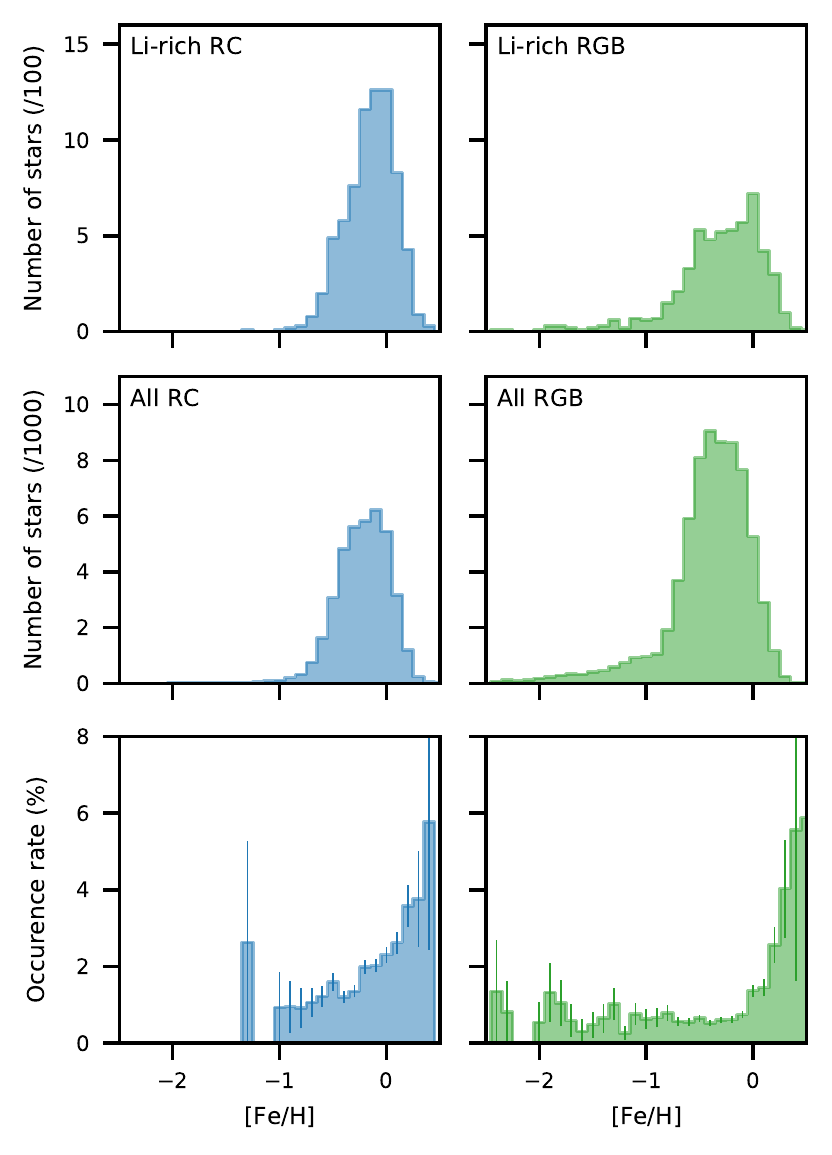}
    \caption{Comparing the occurrence rate of Li-rich giants with changing \feh in the RC (left column) and RGB (right column) cohorts (here just showing the \textsc{bstep} classification method results). In the top panel of each column is the \feh distribution of Li-rich giants from each cohort; the middle panel is the \feh distribution of all stars in each cohort (whether or not they have a measured \ali); and the bottom panel of each column is the occurrence rate with \feh (i.e, the top panel `divided' by the middle panel). The uncertainty on each bin is calculated from Poisson statistics. The occurrence rate of Li-rich giants in both groups rises at high metallicity, but the distributions are quite different --- for the RC stars there is a steady increase of occurrence rate with metallicity, while for RGB stars the occurrence rate is essentially flat below $\feh=0$.}
    \label{fig:feh_li_occurence}
\end{figure}

There has been some tension in the literature between the $<1$~per~cent occurrence rates of Li-rich giants in the low metallicity environments of globular clusters \citep[see e.g.,][]{Kirby2012} and the $>1$~per~cent rate observed in the disk of the Milky Way. Recent works with larger data sets \citep[e.g.,][]{Casey2019,Deepak2020} have quantified this as a more general increase in the occurrence rate of lithium-rich giants with increasing metallicity. In Figure \ref{fig:feh_li_occurence} we consider the occurrence rate of Li-rich giants with metallicity in our RC and RGB cohorts independently. The uncertainty in each bin is calculated from Poisson statistics, i.e., the size of error is estimated to be the fraction of stars that are Li-rich divided by the square root of the number of Li-rich stars). The metallicity distribution of the Li-rich giants is qualitatively similar to the distribution for all giants. As would be expected for the GALAH DR3 sample, which is comprised mainly of Galactic disc stars, the distribution peaks near Solar metallicity. For the RGB sample there is a tail of stars to low metallicity.

The occurrence rate of Li-rich giants in both groups rises at high metallicity, but the distributions are quite different. For the RC stars there is a steady increase of occurrence rate with metallicity, while for RGB stars, the occurrence rate is relatively flat from $-2<\feh<0$, and then increases dramatically for stars with super-Solar metallicity (though with larger error bars due to the smaller number of stars observed at these metallicities). We interpret this increase as a sign of multiple lithium enrichment processes at work, with the dominant mechanism for red clump stars being quite sensitive to stellar metallicity, the dominant process for RGB stars with sub-Solar metallicity being independent of metallicity, and potentially a third lithium enrichment process for metal-rich RGB stars.

\subsection{Other elemental abundances in Li-rich giants}\label{sec:abundances}
The abundances of other elements in Li-rich giants could provide information about the processes by which the lithium abundance of some giants is enhanced. GALAH DR3, which provides abundances for a wide range of elements from a variety of nucleosynthetic pathways, provides an excellent opportunity to explore this.

There are only a few elements in GALAH DR3 that show any obvious differences between the abundance patterns of Li-normal and Li-rich giants. In this section we discuss \cfe, as it was recently highlighted by \citet{Deepak2020} as showing possible correlations with lithium enrichment. Appendix \ref{app:afe} discusses an apparent correlation between the occurrence rate of lithium-rich RC stars and $\upalpha$ enhancement that is driven by the metallicity dependence discussed in Section~\ref{sec:teff_logg_fe_h}.

\begin{figure}
	\includegraphics[width=\columnwidth]{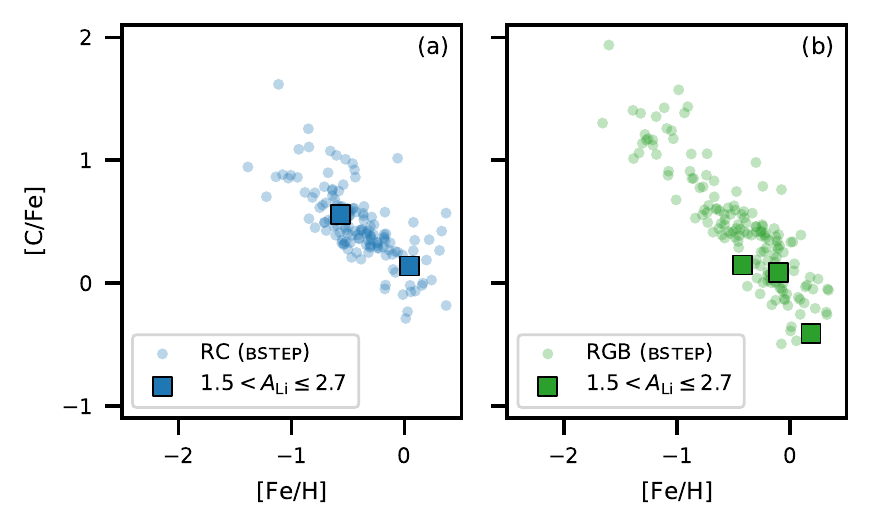}
    \caption{Carbon abundance \cfe for the RC (left) and RGB (right) stars. Carbon can only be measured in HERMES spectra for a small minority of giant stars where it is abundant. The carbon abundance of Li-rich giants is of interest because the two elements are depleted together during first dredge-up, and carbon may also be destroyed by fusion processes at temperatures that can destroy lithium. Only five of the Li-rich giants (all with $1.5<\ali\leq2.7$) have a measured \cfe, and they follow the \ali behaviour of the other stars with measured \cfe.}
    \label{fig:fe_h_c_fe_rc_rgb}
\end{figure}

\citet{Deepak2020} recently explored the other elemental abundances of Li-rich and Li-normal giants using the GALAH DR2 data set. They found that for all the elements available, the only element that showed an appreciable difference between the two populations was carbon. Unfortunately, this result relied upon abundances that had been identified by the GALAH team as unreliable (i.e., the quality flags on the abundances were non-zero). Figure~\ref{fig:fe_h_c_fe_rc_rgb} shows the non-flagged (i.e., reliable) carbon abundances for RC and RGB stars in GALAH DR3. The spectroscopic features of carbon captured in HERMES spectra are quite weak in the bulk of giant stars, and therefore our sensitivity to \cfe is limited. This produces a clear detectability trend that can be seen in the Figure, where lower \cfe abundances are only detected for more metal-rich stars. Only five of the Li-rich giants (all with $1.5<\ali\leq2.7$) have a measured \cfe, and they follow the \ali behaviour of the other stars with measured \cfe.

Interestingly, we do derive reliable (and high) carbon abundances for a number of stars. First dredge-up and subsequent mixing processes typically result in sub-Solar \cfe for giant stars \citep[e.g.,][]{Lagarde2019}. Carbon-richness at this stage in stellar evolution often indicates mass transfer from an asymptotic giant branch (AGB) companion, and may be accompanied by other chemical tags of AGB nucleosynthesis including s-process elements \citep[e.g.,][]{Hansen2016,Karakas2016}. The topic of carbon-enhanced giant stars in GALAH DR3 is outside the scope of this paper, but bears further investigation.

\subsection{Stellar rotation}
\label{sec:rotation}

\begin{figure}
	\includegraphics[width=\columnwidth]{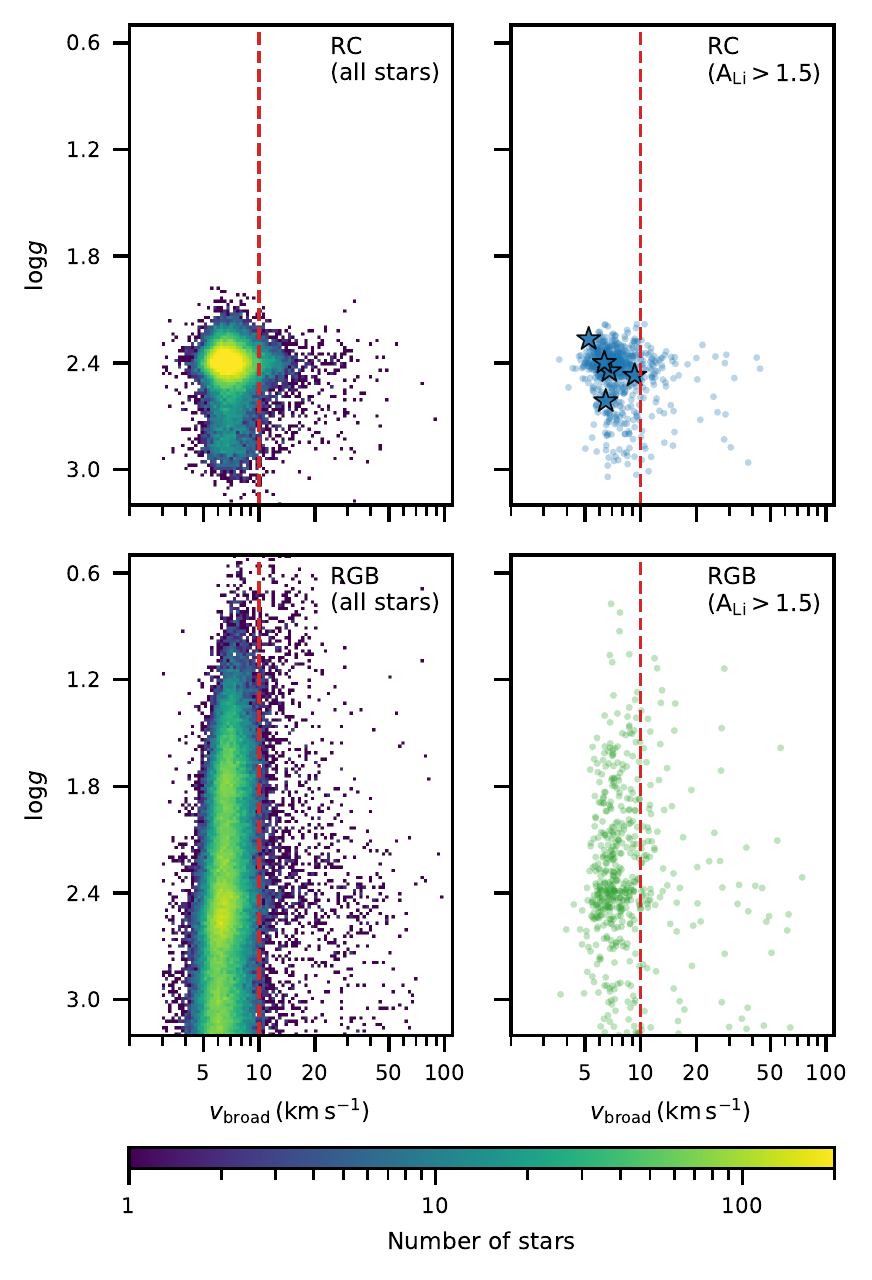}
    \caption{The \vbroad distributions of the red clump (top row) and red giant branch (bottom row) samples. The quantity \vbroad is a broadening term measured from the stellar spectra that encompasses macroturbulence and rotational velocity. The left column shows all giants from the RC and RGB cohorts, while the right column shows just the Li-rich stars ($\ali>1.5$). As in Figure \ref{fig:teff_logg_fe_h_ali_rc_rgb}, in the right column we highlight the stars identified seismically as RC or RGB with star shapes. On all panels, the vertical red dashed line indicates $\vbroad=10$~\kms. We interpret stars to the right of the line as as being rapid rotators. Notably, Li-rich RGB stars are more likely to be rapid rotators than Li-rich RC stars.}
    \label{fig:logg_vbroad_rc_rgb}
\end{figure}

\begin{figure}
	\includegraphics[width=\columnwidth]{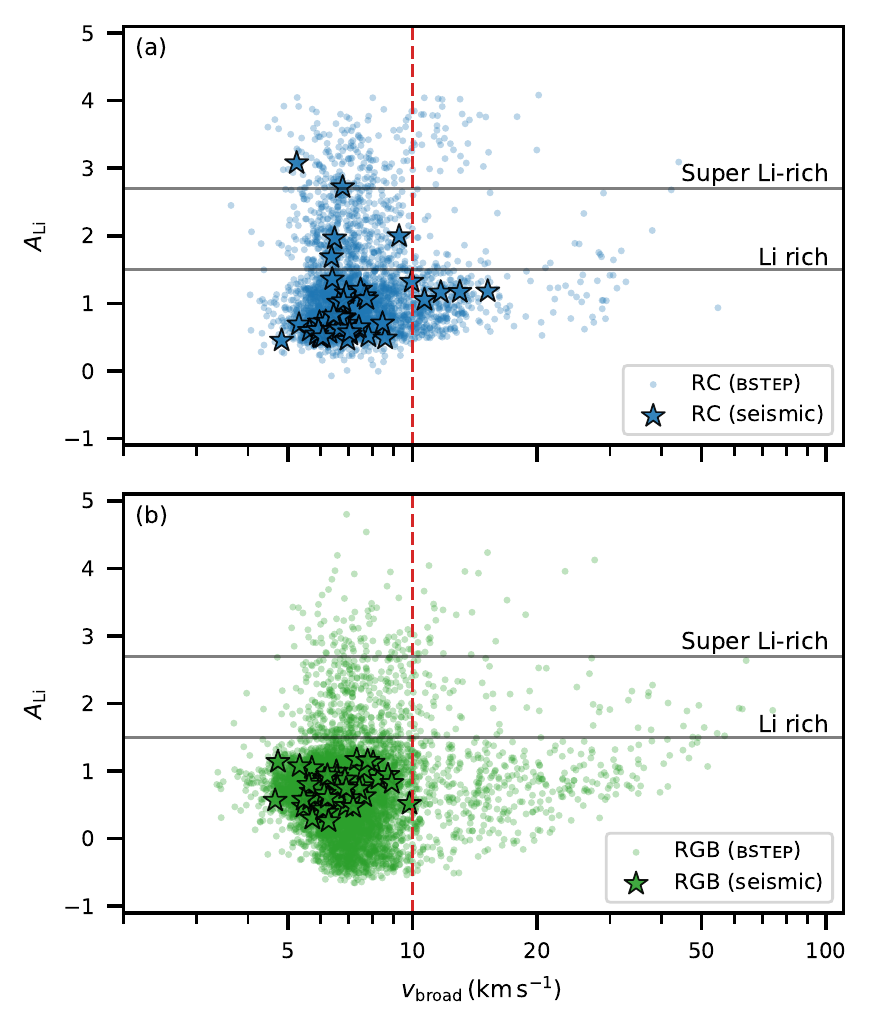}
    \caption{\ali versus \vbroad for the red clump (top panel) and red giant branch (bottom panel) samples. As in Figure \ref{fig:teff_logg_fe_h_ali_rc_rgb}, in each panel we highlight stars asteroseismically classified as RC or RGB. RC stars are not as extended in \vbroad as RGB stars, with most below 20~\kms. The rapid rotators among the Li-rich RC stars are mostly super Li-rich.}
    \label{fig:vbroad_ali_rc_rgb}
\end{figure}

The rotation rates of Li-rich giants are of interest because one of the proposed modes of lithium enhancement is rotationally induced mixing, which can raise the surface lithium abundance via the process described by \citet{Cameron1971}. As part of the spectroscopic analysis in GALAH DR3, an overall spectral broadening parameter \vbroad is calculated, which encompasses macroturbulence and rotational velocity. Typically in RGB stars, the macroturbulence velocity is on the order of 7~\kms \citep{Carney2008}. 

In Figure \ref{fig:logg_vbroad_rc_rgb} we show the distributions of \vbroad with \logg for RC and RGB stars separately. We can see that most stars are found with $5<\vbroad<10~\kms$, so most stars in our sample do not have appreciable rotational velocity.

\begin{table}
\caption{Comparison of how many RC and RGB stars are rapid rotators. Rapid rotators are defined as those stars with \vbroad $>10~\kms$. In the last column, the first percentage is the proportion of all stars of that class that are Li-rich rapid rotators, and the second percentage is the proportion of Li-rich stars in that class that are rapid rotators. RC stars are more likely than RGB stars to be rapid rotators (4.8 per cent versus 3.1 per cent), but the proportion of Li-rich stars that are rapid rotators is higher for RGB stars than RC stars (20 per cent versus 12 per cent).
}
\begin{tabular}{lrrr}
\hline
 & Total stars & Fast rotators & Li-rich fast rotators\\
\hline
All giants & 109340 & 4054 (3.7\%) & 195 (0.18\%; 15\%)\\
RC & 38997 & 1866 (4.8\%) & 89 (0.23\%; 12\%)\\
RGB & 70343 & 2188 (3.1\%) & 106 (0.15\%; 20\%)\\
\hline

\label{tab:fast_rotator_stats}
\end{tabular}
\end{table}

We label stars with \vbroad larger than 10~\kms as ``rapid rotators''. In Table~\ref{tab:fast_rotator_stats} we list the number of stars above this threshold. With this rapid rotation definition, 3.7 per cent (4054/109340) of giants are rapid rotators, which is consistent with the 2 per cent predicted for K giants in the field by \citet{Carlberg2011}. For just the RC stars, the probability to be a rapid rotator $p({\rm RR|RC})=0.048$ (1866/38997), while for the RGB stars $p({\rm RR|RGB})=0.031$ (2188/70343); that is, RC stars are about 1.5 times as likely as RGB stars to be rapid rotators. However, considering only Li-rich stars, the probability to be a rapid rotator is very different, with $p(\textrm{RR|Li-rich,RC})=0.122$ (89/726) and $p(\textrm{RR|Li-rich,RGB})=0.20$ (106/536). Hence, for the Li-rich stars, the RGB stars are nearly twice as likely as RC stars to be rapid rotators.

Considering Li richness for a given evolutionary phase and state of rotation (rapidly rotating versus non-rapidly rotating) provides more insight into the above result, since it allows us to directly investigate whether Li richness is correlated with rapid rotation, and whether that connection is stronger for RC versus RGB stars. The probability to be Li-rich for a non-rapidly-rotating star is $p(\textrm{Li-rich} | \textrm{NRR,RC})=0.017$ (637/37131) for an RC star, and $p(\textrm{Li-rich} | \textrm{NRR,RGB}) = 0.0063$ (430/68155) for an RGB star. However, the probability to be Li-rich for a rapidly rotating star is much higher and is equal for both RGB and RC stars: $p(\textrm{Li-rich} | \textrm{RR,RC}) =0.048$ (89/1866), and $p(\textrm{Li-rich} | \textrm{RR,RGB}) = 0.048$ (106/2188). This indicates that rapid rotation is a mechanism for Li enrichment, for both RGB and RC stars.

In Figure~\ref{fig:vbroad_ali_rc_rgb} we explore the relationship between stellar rotation and \ali. Here we plot the \vbroad, noting those stars with values $>10~\kms$ as rapid rotators. For both the RC and RGB populations, the Li-normal stars ($\ali<1.5$) show a range in this \vbroad, though the RC stars are more restricted, with only a few above 20~\kms. Curiously, the rapid rotators in the RC group are quite skewed toward being super Li-rich, with only a few RC stars with $1.5<\ali<2.7$ being rapid rotators.

\subsection{Binarity}\label{sec:binarity}
Recent works have presented two models involving binary stars to explain lithium enrichment on the red clump. \citet{Casey2019} used a large set of Li-rich giants from the LAMOST survey to argue that the distribution of Li-rich giants in evolutionary phase requires enrichment processes operating both at random points on the first-ascent giant branch and at the helium flash, based on the lithium depletion timescale in stellar atmospheres. Their proposed mechanism for lithium production in red clump stars is tidal spin-up from a binary companion driving internal mixing and therefore lithium production via the \citet{Cameron1971} mechanism. In contrast, \citet{Zhang2020} proposed that Li-rich RC stars are the result of mergers in RGB-white dwarf binary systems.

These hypotheses are in principle testable by searching for binary companions (or lack thereof) in Li-rich giants through variability in radial velocity, photometry and astrometry. Unfortunately, as we discuss in this subsection, we are not able to draw any significant conclusions on binarity in our data set from the available data. Recent work by \citet{Traven2020} analyzed the GALAH survey spectra to identify spectroscopic binary stars, and found a handful of RC binaries, including a few that were lithium-rich. Followup work to re-derive the lithium abundances of the individual stars could be quite useful in understanding how important binary RC stars are in the family of lithium-rich giants.

\citet{Casey2019} used a binary comprised of a 1.5 $M_{\odot}$ giant and a 1.0 $M_{\odot}$ dwarf as an example system, and calculated the range of orbital periods capable of producing observable lithium enrichment in the giant. The minimum orbital period of 279 days was set by the need to avoid mass transfer through Roche lobe overflow. Such an orbit would have observable spectroscopic and photometric signatures, which would be less pronounced for the longer orbits that would also be sufficient to produce lithium enrichment for this particular pair of stars. For this system, we would expect radial velocity variations of $\sim30~\kms$ or less \citep[as modelled by \textsc{ellc};][]{Maxted2016}. This is well within the precision of HERMES but we do not have the necessary observational cadence to confidently identify these periodic variations. \citet{Price-Whelan2020b} identified 19,635 candidate binaries in the APOGEE survey based on radial velocity variations between multiple observations. Of these, 53 are in our giant star data set, including 15 of our RC stars --- one of which is Li-rich. It has only three observations with APOGEE, indicating an RV range of $\sim50~\kms$. 

In terms of photometric signatures of binarity in the form of transits, assuming random orbital inclinations and observed orbital period distributions \citep[e.g.,][]{Raghavan2010}, $2.5\pm0.5$ per cent of our Li-rich RC stars should be in eclipsing binary systems. The secondary star will block about 1 per cent of the disk of the primary with an eclipse period of $\sim3$ days. This does not require a very high cadence in photometric monitoring, but none of our Li-rich RC stars have so far had light curves in ASAS-SN \citep{Kochanek2017} or \tess\ \citep{Ricker2015} that show evidence for eclipses.

The astrometry from \gaia is precise enough to measure the motion of the photocentre of some binary systems (depending on heliocentric distance and the binary properties). The model assumption underlying the data processing for \gaia DR2 is that sources are single objects, and if there are photocentre shifts in binary systems these are interpreted as larger than expected astrometric errors. Work by \citet{Penoyre2020} and \citet{Belokurov2020} have used large values of the Renormalized Unit Weight Error (\texttt{RUWE}), an astrometric error metric reported by \gaia, to identify possible binaries. Our data set excludes stars with large \texttt{RUWE} by construction (it is part of \texttt{flag\_sp}), because parallax is an important prior in our stellar parameter determination. 

\subsection{Infrared excess}\label{sec:ir_excess}
\begin{figure}
	\includegraphics[width=\columnwidth]{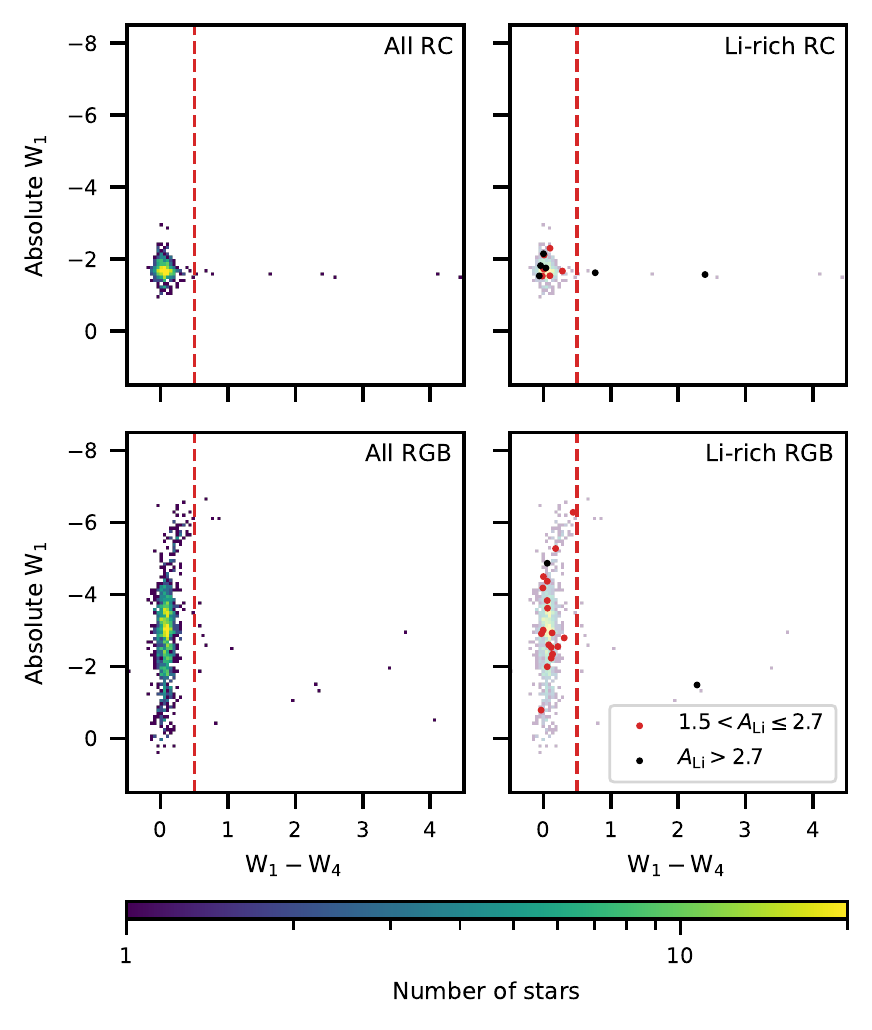}
    \caption{The infrared colour-magnitude diagram for stars in our data set with reliable \wise\ photometry, showing (top row) RC stars, (bottom row) RGB stars, and (left column) all stars in each sample and (right column) Li-rich stars highlighted with red and black circles. Stars to the right of vertical line at $W_{1}-W_{4}=0.5$ have an infrared excess. Although most Li-rich stars do not have an IR excess, the proportion of IR excess stars that are Li-rich (3/25; 12 per cent) is much higher than the proportion of the general population of giants that are Li-rich (33/1862; 1.8 per cent).
    }
    \label{fig:wise_rc_rgb_A}
\end{figure}

A fraction of Li-rich RGB stars have been reported to have excess flux in their spectral energy distribution in the infrared, which has been postulated as a sign of a physical connection between lithium production and mass loss. This can be seen in photometry from the \textit{IRAS} satellite \citep[e.g.,][]{Fekel1998} and from \wise, although the latter's wavelength coverage does not reach as far into the infrared. A close investigation by \citet{Rebull2015} found that the majority of lithium-rich giants in the literature with reported infrared excesses were artifacts in the \wise\ catalog or cases of source confusion. However, they do confirm some as real cases of infrared excess, and they do find that the stars with the largest infrared excess are lithium-rich K giants.

Within our data set, there are only 1862 giants with clean \wise\ detections; that is, they have the \texttt{cc\_flags} confusion flag set to \texttt{0000} and the \texttt{ph\_qual} photometric quality flag set to \texttt{A} for $\mathrm{W}_1$ and $\mathrm{W}_4$. Figure \ref{fig:wise_rc_rgb_A} shows an infrared colour-magnitude diagram for these stars, divided in the same way as Figure \ref{fig:logg_vbroad_rc_rgb} with red clump stars in the upper panels, red giant branch stars in the lower panels, all stars in the left panels, and the lithium-rich stars highlighted in the right panels. There are 549 RC stars (13 of which are Li-rich) and 1313 RGB stars (20 Li-rich). Overall only a few stars have large infrared colours (defined as $W_{1}-W_{4}>0.5$): 8 of the RC stars and 17 of the RGB stars. Of the 33 Li-rich giants with useful \wise\ photometry, three have large infrared colours. All are super Li-rich, with two classified as RC stars. It is clear that infrared excess is not a requirement for lithium enrichment, but stars with infrared excess are more likely to be lithium-rich than stars without infrared excess. Since only a small subsample of our data has reliable \wise\ photometry, we cannot draw strong conclusions about how mass loss, lithium enrichment and evolutionary phase are related from this data set.

\subsection{Super lithium-rich stars}
\label{sec:superlirich}
In our discussion and figures so far we have been making a distinction between lithium-rich stars, which have $\ali>1.5$, and super lithium-rich stars, which sit above the primordial abundance of $\ali=2.7$. This is an important distinction for any model of lithium enrichment that involves adding pristine gas to an evolved star to raise its abundance, because that process can only raise a star's abundance toward the initial abundance in that gas but could not exceed it. 

For lithium enrichment models that require the production of lithium within a star, this distinction is less important, but the question of how to produce the observed amount of lithium is more difficult to answer at higher abundance. \citet{Yan2018} attempt to model the process for lithium production in the star TYC~429-2097-1, which has an abundance of $\ali=4.5$. They find that meridional circulation at the RGB bump, the current evolutionary stage of this star, is capable of producing more than the observed amount of lithium. The most lithium-rich giant in our data set 
has $\ali=4.8$\footnote{See Section \ref{sec:li_abundances} for 
a caution about the abundances of very lithium-rich stars}, but as a luminous giant (see Table \ref{tab:params} and Figure \ref{fig:teff_logg_fe_h_ali_rc_rgb}) it does not fit within their model.

Super lithium-rich red clump stars in our data set have some interesting properties. While the majority of our super lithium-rich stars are in the red clump phase (Section \ref{sec:phase}), very few of them are on the secondary red clump, indicating some difference between pRC and sRC stars in terms of their ability to produce or retain large amounts of lithium. Expanding the data set of precise asteroseismic masses for red clump stars would be extremely useful for future work investigating the source of this difference. While the super Li-rich red clump stars are the most rapidly rotating (Fig. \ref{fig:logg_vbroad_rc_rgb}) and the most metal-rich (Fig. \ref{fig:fe_h_alpha_fe_rc_rgb}), these two groups of super Li-rich red clump stars are not the same set of stars: the most metal-rich RC stars are not rapidly rotating.

\subsection{Summary of observational phenomenology}
In this study we explore the properties of 1262 evolved stars from the GALAH and \Ktwo\ surveys with elevated photospheric abundances of lithium $(\ali>1.5)$. We find these main behaviours in the data set:
\begin{enumerate}
    \item Red clump stars are 2.5 times as likely to be lithium-rich as red giant branch stars (Section \ref{sec:phase}).
    \item The less massive primary red clump stars are 1.5 times as likely to be super lithium-rich as the more massive secondary red clump stars (Section \ref{sec:phase}).
    \item The occurrence rate of lithium-rich giants with metallicity is markedly different for red clump and red giant branch populations: it increases steadily with metallicity in red clump stars, but it is essentially constant in red giant branch stars below Solar metallicity and increases sharply thereafter (Section \ref{sec:teff_logg_fe_h}).
    \item The probability for a star to be lithium-rich is almost five times as high for rapidly rotating stars as for slowly rotating stars (Section \ref{sec:rotation}). 
    \item Lithium-rich red giant branch stars are 1.5 times as likely as lithium-rich red clump stars to be rapidly rotating, but the probability for a rapidly rotating star to be lithium-rich is the same for red giant branch and red clump stars (Section \ref{sec:rotation}).
    \item The majority of lithium-rich giants do not have an infrared excess, but the stars with infrared excess are five times as likely to be lithium-rich as the stars with no infrared excess (Section \ref{sec:ir_excess}).
    \item Rapidly rotating lithium-rich red clump stars tend to be super lithium-rich (Section \ref{sec:superlirich}).
\end{enumerate}

A number of previous studies have discussed the fraction of lithium-rich giants that are in the core helium burning phase versus the first-ascent red giant branch phase. This distribution across evolutionary state is a key piece of information for understanding the origin of lithium enrichment, and it underlines the need for models that address stars from first dredge-up through to core helium burning. It was difficult to generalise from the data sets presented in earlier studies to the overall population of lithium-rich giants, since selection effects could drive them significantly toward one particular type of star (eg, luminous red giants, as in \citealt{Gonzalez2009} and \citealt{Lebzelter2012}, or stars on the red clump, as in \citealt{Kumar2011} and \citealt{Singh2019a}). Large-sample studies such as \citet{Deepak2019}, \citet{Casey2019}, and \citet{Yan2021} found that 68\%, 80\%, and 86\%, respectively, of lithium-rich giants belong to the red clump. 

In this study we follow the lead of \citet{Casey2019} and \citet{Deepak2020}, and look at the probability for a star in a given evolutionary phase to be lithium-rich, rather than the probability for a lithium-rich star to be in a given evolutionary phase. The overall occurrence rate is known to lie around 1\%, but subdividing the data set into red clump versus red giant branch stars and looking at the occurrence rate as a function of metallicity brings out some interesting behaviours. We recover the changing occurrence rate of lithium-rich giants with metallicity noted by \citet{Casey2019} and \citet{Deepak2020}. We do find a smaller fraction of our Li-rich giants to be RC stars than in \citet{Casey2019}, 64 per cent versus their $80^{+7}_{-6}$ per cent. We attribute this difference to a number of factors. The spectroscopic inference method for identifying red clump stars in \citet{Casey2019} has a precision of 97\%, while our isochrone method is 82\% precise. Since the red clump stars are more likely to be lithium-rich than the red giant branch stars, any misclassification between the two groups will tend to reduce the fraction of lithium-rich red clump stars. With the different spectral resolution and lithium abundance determination methods between the two projects, there are likely different detectability limits and abundance accuracies that could move stars near \ali$=1.5$ across the lithium-rich dividing line. Finally, LAMOST and GALAH have very different selection functions, which could easily result in different proportions of red clump and red giant branch stars in the observed data sets. 

Super lithium-rich stars in the literature (those with A(Li)$\geq 3.3$) do have a tendency to be found in the red clump phase, though the difference we see in the occurrence rate between primary and secondary red clump stars and the correlation between rapid rotation and very high lithium abundance in red clump stars are not noted in previous studies. Studies with asteroseismic classifications \citep{Du2020, Deepak2019, Singh2019a, Singh2019b, Kumar2018} identify 53 super lithium-rich stars belonging to the red clump, with an additional three in \citet{Deepak2019} that appear to be early AGB stars. In studies without asteroseismic data, \citet{Monaco2014} identifies one super lithium-rich red clump star in the open cluster Trumpler 5, four of the seven super lithium-rich stars in \citet{Smiljanic2018} have \teff and \logg consistent with the red clump, and three of six super lithium-rich stars in \citet{MS13} have stellar parameters consistent with the red clump, with the remaining super lithium-rich stars being two subgiants and an AGB star. The studies of \citet{Holanda2020b}, \citet{Zhou2018}, and \citet{Lyubimkov2015} each report one super lithium-rich ``cool giant'' or ``K giant'', but in all cases the stellar parameters are consistent with the red clump.

We do not replicate the upper limit on RGB star lithium abundances reported in LAMOST data by \citet{Yan2021} in our full data set, but we do see a similar pattern in the asteroseismically confirmed red giant branch stars, none of which are lithium-rich (Figure \ref{fig:teff_logg_fe_h_ali_rc_rgb}, lower middle panel). The asteroseismic classifications from \citet{Hon2018} only apply for stars on the lower giant branch, since the oscillation frequencies for the more luminous giants are too low to be well determined from \Ktwo\ data. The precision of \textsc{bstep} classifications for stars near the luminosity of the red clump is 82\%, so there is a possibility that the lithium-rich RGB stars without asteroseismic data at that luminosity are misclassified RC stars. However, the classification of our bright red giants is unambiguous even without asteroseismology, and we do find stars on the upper giant branch with enhanced lithium abundances, including some above the primordial level. If this enrichment limit is real, is suggests that the enrichment mechanism for RGB stars near the RGB bump may somehow be constrained in the amount of lithium it can produce, while RGB stars enriched later can receive larger enhancements. Another possible interpretation is that stars that have recently undergone lithium enrichment have asteroseismic power spectra that are more difficult for the neural network method of \citet{Hon2018} to analyse. It would be useful to check the known lithium-rich stars with asteroseismic classifications from \textit{Kepler} to see if this distinction between lower and upper RGB stars is also found in that data set.

Both rapid rotation and infrared excess show some connection to lithium enrichment, in that stars with those properties are more likely to be lithium-rich than stars without them. However, the majority of lithium-rich stars do not exhibit either of these features. As was also observed by \citet{Zhou2019}, the Li-rich stars with infrared excess are not the same Li-rich stars as those that are rapid rotators.

\section{The origins of lithium enrichment in evolved stars} \label{sec:discussion}
The broad strokes of the proposed explanations for how a small fraction of evolved stars have come to be enriched in lithium have not changed substantially since the first lithium-rich red giant was identified by \citet{ws82}. Models invoke either some external reservoir of lithium (ingestion of a planet or sub-stellar companion; or mass transfer from an AGB companion) or some internal production channel (internal mixing driven by the RGB bump phase, the He flash, or rotation). Binary interactions have also been proposed as a way of inducing internal production, either through the merger of a red giant branch star and a white dwarf, or a tidal interaction with a binary companion that spins up a red clump star.

The prevailing opinion in recent observational studies is that there are likely multiple processes responsible for enhanced lithium abundances observed in stars near the red giant branch bump, luminous red giant branch stars, and red clump stars. Certainly, some of the proposed mechanisms, such as planet engulfment or extra mixing at the red giant branch bump, are not viable as the only source for lithium enrichment across such a heterogenous group of stars. This study also finds that lithium-rich giants are a diverse population, which is consistent with the idea of multiple mechanisms, but does not absolutely require it.



The correspondence in red clump stars between rapid rotation and the highest levels of lithium enrichment implies quite strongly that rotationally driven mixing processes are capable of causing lithium production, and the relative lack of secondary red clump stars with very high levels of lithium enrichment suggests that they are less likely to experience large amounts of lithium enrichment, or more likely to re-deplete it during their helium burning lifetime.

Although spectroscopy allows us to measure surface rotation\footnote{We note that in this study we are measuring the combination of rotation and macrotuburbulence}, the rotation profile throughout the star is a key factor in the strength of rotationally induced mixing. Relatively rapid surface rotation can be observed in some giant stars, but the cores and atmospheres of post-main sequence stars are mechanically decoupled, and the surface rotation does not provide sufficient information on their internal rotation. 
Asteroseismology presents an opportunity to measure the internal rotation of evolved stars that have rotational axes aligned appropriately to the line of sight \citep[e.g.,][]{Mosser2012,Deheuvels2012}. Studies of additional Li-rich giants in the Kepler field and the TESS continuous viewing zone would be very helpful in understanding whether there is a connection between core rotation and lithium enrichment.

For red giant branch stars, we find that there is a concentration of lithium-rich stars near the luminosity of the red giant branch bump, which can be explained as a result of internal mixing triggered by a change in internal structure. We also find lithium-rich stars at all luminosities and metallicities on the red giant branch. The occurrence rate for lithium-rich red giant branch stars is essentially constant for all sub-Solar metallicities and dramatically higher at super-Solar metallicity. This indicates that the efficiencies of the processes that produce and then re-deplete lithium in these stars may be dependent on metallicity, or that there may be an additional lithium production process at work in metal-rich red giant branch stars. It is unclear how compatible a metallicity-dependent lithium enrichment process on the red giant branch is with planet engulfment models, since the planet occurrence rate for gas giants is not a straightforward function of metallicity. Further abundance studies focused on boron and beryllium could shed some light on this problem, since they have similar burning temperatures to lithium and should be also be enriched during planet engulfment \cite[e.g., see initial work by][]{Drake2017,Carlberg2018}. Planet engulfment is expected to result in a different ratio of $^6$Li/$^7$Li compared to internal lithium production processes, but measuring this will be observationally challenging \citep{Aguilera-Gomez2020}.

As in previous studies, the majority of our lithium-rich giants are red clump stars, and this requires a lithium enrichment process triggered at, or after, the helium flash. The fact that the occurrence rate for red clump stars rises steadily with increasing metallicity may be a result of the fact that the time spent on the red clump is longer at higher metallicity, or it may reflect a more effective mixing-driven lithium production in high metallicity giant stars because of their less compressed interior structure, or it may result from a higher binary fraction (with the correct mass ratio and orbital separation) at higher metallicity. 
With the present data set we cannot comment directly on binary interactions as the driver for internal mixing, but this is an avenue for future work that may clarify the situation significantly.

Recent work by \citet{Kumar2020} indicates that lithium enrichment at the helium flash may be compulsory, given the lack of very lithium-depleted red clump stars in their data set. This provides an intriguing and testable prediction for red clump stars as a population, and may provide new insights into the internal rearrangement that happens as a result of the helium flash. Both detailed modeling of stellar structure and evolution, and population synthesis modeling similar to \citet{Casey2019}, will be essential for evaluating the mechanisms for lithium enrichment and re-depletion, and for understanding their dependence on stellar mass, composition, and rotational velocity.

\section*{Acknowledgements}

SLM, JDS and AGB acknowledge support from the UNSW Scientia Fellowship program and from the Australian Research Council through Discovery Project grant DP180101791.
SB acknowledges funds from the Alexander von Humboldt Foundation in the framework of the Sofja Kovalevskaja Award endowed by the Federal Ministry of Education and Research.
Y.S.T. is grateful to be supported by the NASA Hubble Fellowship grant HST-HF2-51425.001 awarded by the Space Telescope Science Institute
Parts of this research were conducted by the Australian Research Council Centre of Excellence for All Sky Astrophysics in 3 Dimensions (ASTRO 3D), through project number CE170100013.

This work is based on data acquired through the Anglo-Australian Telescope, under programmes: The GALAH survey (A/2013B/13, A/2014A/25, A/2015A/19, A2017A/18); The \Ktwo-HERMES \Ktwo\ follow-up program (A/2015A/03, A/2015B/19, A/2016A/22, A/2016B/12, A/2017A/14); 
Accurate physical parameters of \textit{Kepler} \Ktwo\ planet search targets (A/2015B/01); Planets in clusters with \Ktwo\ (S/2015A/012). We acknowledge the traditional owners of the land on which the AAT stands, the Gamilaraay people, and pay our respects to elders past and present.

This work has made use of data from the European Space Agency (ESA) mission \gaia (\url{https://www.cosmos.esa.int/gaia}), processed by the \gaia Data Processing and Analysis Consortium (DPAC, \url{https://www.cosmos.esa.int/web/gaia/dpac/consortium}). Funding for the DPAC has been provided by national institutions, in particular the institutions participating in the \gaia Multilateral Agreement.

This paper includes data collected by the \Ktwo\ mission. Funding for the \Ktwo\ mission is provided by the NASA Science Mission directorate.

This publication makes use of data products from the Wide-field Infrared Survey Explorer, which is a joint project of the University of California, Los Angeles, and the Jet Propulsion Laboratory/California Institute of Technology, funded by the National Aeronautics and Space Administration.

This project was developed in part at the Expanding the Science of TESS meeting, which took place in 2020 February at the University of Sydney. 

The following software and programming languages made this research possible:
\textsc{astropy} \citep[v4.0.1;][]{astropy18};
\textsc{ellc} \citep[1.8.5;][]{Maxted2016};
\textsc{gala} \citep[v1.1;][]{galasw};
\textsc{h5py} (v2.10.0);
\textsc{matplotlib} \citep[v3.1.3;][]{Hunter2007,Caswell2020};
\textsc{mpl\_scatter\_density};
\textsc{python} (v3.8.3);
\textsc{seaborn} \cite[v0.10.1;][]{Waskom2020};
\textsc{scipy} \citep[v1.4.1;][]{SciPy1.0Contributors2020};
\textsc{topcat} \citep[v4.7][]{Taylor2005,Taylor2006}.

\section{Data availability}
The GALAH DR3 catalogue, several value-added catalogues, and all HERMES spectra of the sources are available for download via the Data Central service at \url{datacentral.org.au}. The accompanying documentation can be found at \url{docs.datacentral.org.au/galah}, and a full description of the data release is given in \citet{Buder2020}.

\bibliographystyle{mnras}

\appendix
\section{Alpha enhancement and lithium-rich giants}
\label{app:afe}
\begin{figure}
	\includegraphics[width=\columnwidth]{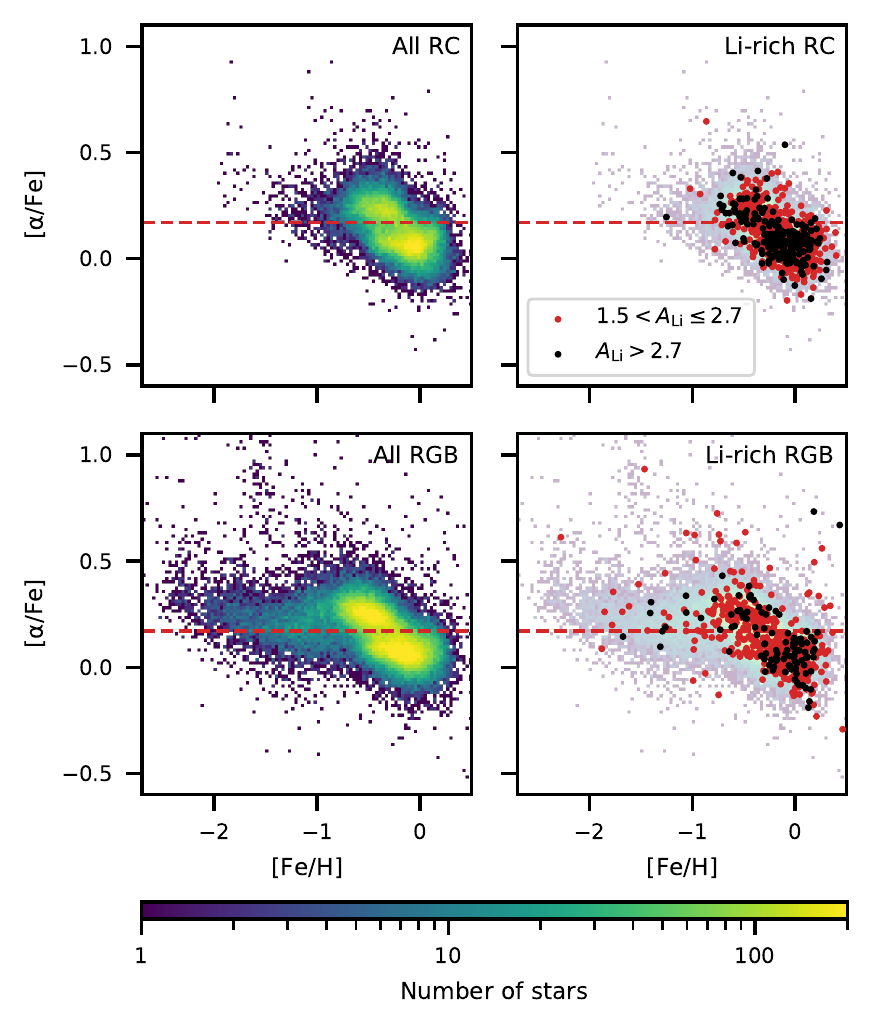}
    \caption{Comparison of the \alphafe abundances of the RC stars (upper panels) and RGB stars (lower panels). In the right column we indicate the Li-rich and super Li-rich giants with red and black dots, respectively. The majority of super Li-rich RC stars belong to the $\upalpha$-low ``thin disc'' population.}
    \label{fig:fe_h_alpha_fe_rc_rgb}
\end{figure}
For GALAH DR3, \alphafe is the error-weighted combination of the abundances determined from selected Mg, Si, Ca, and Ti lines. Figure~\ref{fig:fe_h_alpha_fe_rc_rgb} shows our data set in the \alphafe versus \feh plane, divided into RC stars (upper panels) and RGB stars (lower panels), with the number density of stars stars shown in the left panels and the Li-rich and super Li-rich stars highlighted in the right panels with red and black dots, respectively. It is well known that this particular abundance plane records Type II and Type Ia supernova contributions to Galactic chemical evolution over time \citep[e.g.,][]{Venn2004}. Old stars in the halo tend to have low metallicity and high $\upalpha$ enhancement, while above \feh$\approx -1$ the stars divide into the thick disc, where moderately old stars are still $\upalpha$-enriched, and the thin disc, where stars formed up to the present day have no $\upalpha$ enhancement. Each spectroscopic analysis has its own abundance scale, so different surveys find this \alphafe--\feh bifurcation at a slightly different level.

For our purposes stars are labelled as ``$\upalpha$-low'' if they have \alphafe<$0.17$ and ``$\upalpha$-high'' otherwise. At a glance, it appears that Li-rich (red dots) or super Li-rich (black dots) RC stars appear to be preferentially located at low \alphafe values and high \feh, but thin disc stars are dominant in the overall sample, and this may be giving a false impression. To explore this further, we calculate $p(\textrm{Li-rich}|\upalpha\textrm{-high,RC})$, the probability that an $\upalpha$-high RC star is Li-rich, and $p(\textrm{Li-rich}|\upalpha\textrm{-low,RC})$, the probability that an $\upalpha$-low RC star is Li-rich. We find that $\upalpha$-low RC stars are somewhat more likely to be Li-rich than $\upalpha$-high RC stars are: $p(\textrm{Li-rich}|\upalpha\textrm{-high,RC})=0.014$ (185/13597), while $p(\textrm{Li-rich}|\upalpha\textrm{-low,RC})=0.021$ (540/25357). For super Li-rich stars, the probabilities are$p(\textrm{Li-super}|\upalpha\textrm{-high,RC})=0.004$ (50/13597), while $p(\textrm{Li-super}|\upalpha\textrm{-low,RC})=0.006$ (163/25357). 
This does indicate a difference in the prevalence of lithium-rich giants in the thin and thick discs. However, we hesitate to make strong or quantitative statements about this, given the uncertainties introduced by our classification of stars as RC or RGB, and our rough division of the sample at \alphafe$=0.17$.

The apparent affinity between high $\ali$ and low $\alphafe$ does not happen because the lithium enrichment process destroys $\upalpha$ elements, or is hampered by their presence. Rather, it is related to the properties of the Galactic components that are captured in our observational sample, similar to the effect seen by \citet{Ramirez2012}. For stars in the Milky Way disc, including the majority of our RC stars, $\alphafe$ is anticorrelated with metallicity. As shown in \autoref{fig:feh_li_occurence}, the occurrence rate of Li-rich RC stars increases with metallicity, with the result that there are more Li-rich stars in the $\upalpha$-low population.

In contrast to RC stars, RGB stars in both the Li-normal and Li-rich subsets extend to lower $\feh$ and higher $\alphafe$, and the distributions of the Li-normal and Li-rich RGB stars in Figure \ref{fig:fe_h_alpha_fe_rc_rgb} look quite similar to each other, with perhaps an excess of Li-rich stars at $\feh >0$. This is consistent with the occurrence rate of Li-rich RGB stars shown in \autoref{fig:feh_li_occurence}, which is flat for $\feh< 0$ but rises thereafter. RGB stars have higher luminosities than RC stars, and as a consequence in a magnitude limited survey like ours they are drawn from a larger volume including more of the thick disk and the halo. This, in addition to the dependence of occurrence rate on metallicity, skews the distribution of Li-rich RC stars toward low $\alphafe$ and high metallicity relative to RGB stars. 

\section{Spatial and orbital properties}
\label{app:orbits}

\begin{figure}
	\includegraphics[width=\columnwidth]{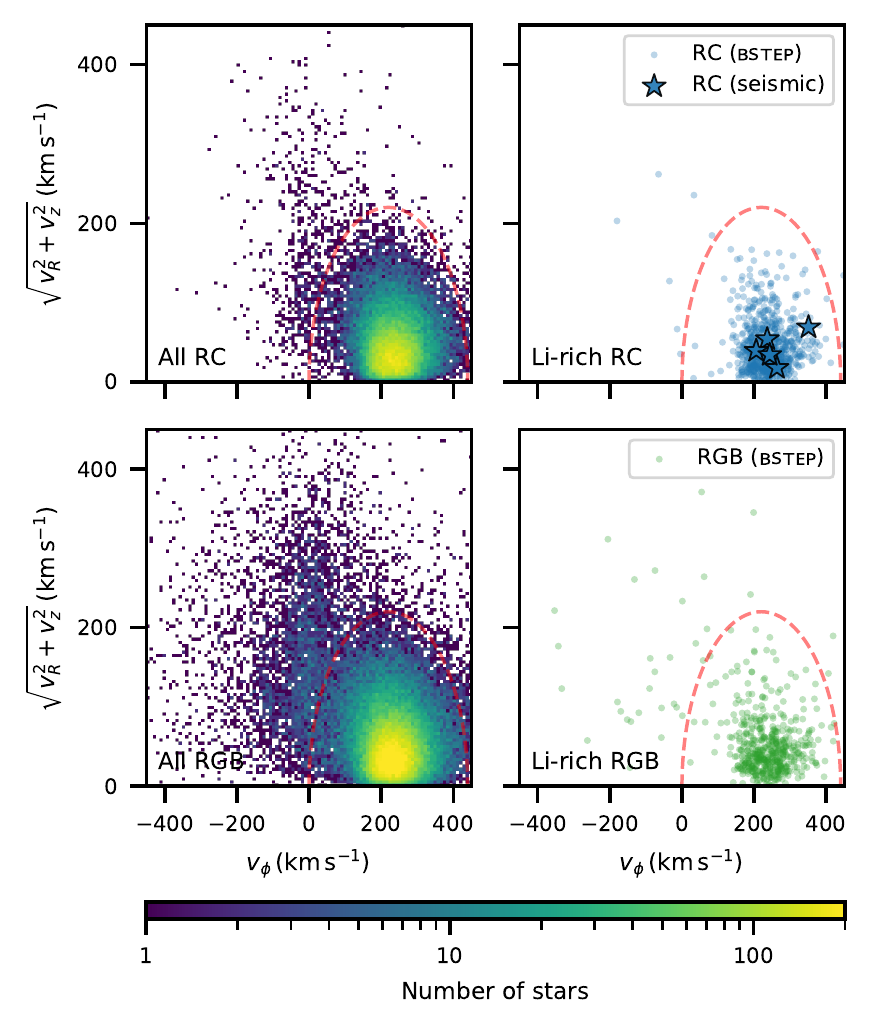}
    \caption{Galactic orbital velocities presented in the form of Toomre diagrams for (left) all stars in each sample and (right) just Li-rich stars. The top row shows RC stars; the bottom row is RGB stars. The red dashed circle on all panels indicates the region of this velocity space within which stars have disk-like orbits; stars outside of the circle have orbital velocities typical for halo stars. As expected for the GALAH survey selection, most of the stars observed have disk-like orbits, in the lithium-normal and lithium-rich groups.}
    \label{fig:toomre_rc_rgb}
\end{figure}

\begin{figure}
    	\includegraphics[width=\columnwidth]{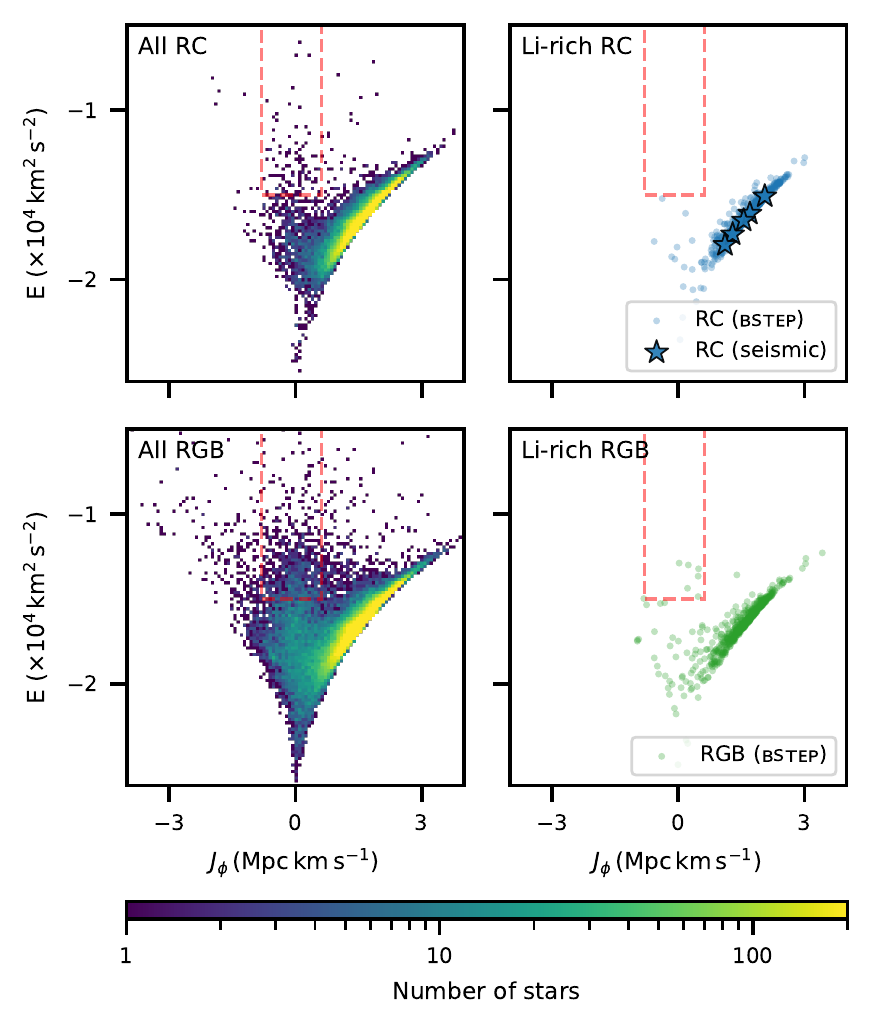}
    \caption{The $J_{\phi}$ vs orbital energy distribution for our stars, with panels arranged the same as in Figure~\ref{fig:toomre_rc_rgb}. As also shown in Figure~\ref{fig:toomre_rc_rgb}, the majority of our Li-rich giants have disk-like orbits --- i.e., they are largely concentrated on the right edge of the envelope with low eccentricity and in-plane motions ($[-2<E<-1]\,10^5\,\mathrm{km}^2\,\mathrm{s}^{-2}$ and $[0<L_z<4]\,\mathrm{Mpc}\,\kms$). In all panels, the red-dashed rectangle indicates the region of parameter space where stars from the \gaia-Enceladus merger event are found \citep{Koppelman2018}. A handful of our Li-rich RGB stars fall into this box, and a more detailed investigation to chemically and kinematically tag these stars to the \gaia-Enceladus merger event is planned.}
    \label{fig:lz_energy_rc_rgb}
\end{figure}

The kinematic and chemical properties of stars in the Milky Way carry information about their origins and subsequent dynamical evolution. Kinematic substructures and mismatches between the data and a smooth distribution in an age-abundance-kinematics space can indicate important events like minor mergers \citep[e.g.,][]{Koppelman2018,Myeong2019,Borsato2020} and radial migration \citep[e.g.,][]{Buder2019,Hayden2020}.

Here we use orbital properties calculated for the stars as described in \citet{Buder2020}. This used GALAH DR3 radial velocities and distances from the age and mass value-added catalog, which primarily incorporated distances found by the Bayesian Stellar Parameters estimator \citep[\textsc{bstep}; described in][]{Sharma2018}, which calculates distance simultaneously with mass, age, and reddening. The orbits are calculated using \textsc{galpy}, with the \textsc{McMillan2017} potential \citep{McMillan2017} and the values $R_\mathrm{GC}=8.21$~kpc and $v_\mathrm{circular}=233.1~\kms$. We set $(U,V,W)_\mathrm{\odot}=(11.1, 15.17, 7.25)~\kms$ in keeping with \citet{Reid2004} and \citet{Schonrich2010}.

The GALAH survey primarily samples the disk of the Milky Way, with only one per cent of the observed stars belonging to the halo \citep{deSilva15}. The locations and metallicities of the Li-rich giants, both the RC and RGB stars, are consistent with them being mainly a disk population. This aligns well with our existing understanding of lithium-rich giants as ordinary stars.

Figure~\ref{fig:toomre_rc_rgb} shows Galactocentric rotational velocity versus perpendicular velocity (i.e., a Toomre diagram). The red circle shows the point where the total velocity relative to the Local Standard of Rest is 220~\kms, which is a canonical division between the disk and the halo \citep{Helmi2008}. In our data set, red clump stars as a whole are more likely to be on disk-like orbits than halo-like orbits, and very few of the Li-rich RC stars are on halo-like orbits. The majority of RGB stars in our data set also orbit in the disk, but both Li-normal and Li-rich RGB stars are more likely to be on halo-like orbits than RC stars are. This is consistent with the different volumes these two sets of stars sample in a magnitude-limited survey like GALAH.

We can further investigate the halo RGB stars in Figure~\ref{fig:lz_energy_rc_rgb}, which shows the orbital energy $E$ and the azimuthal action $J_{\phi}$ ($\equiv L_{z}$; vertical angular momentum). This is a coordinate space in which prograde orbits are on the right side of the plot and retrograde orbits are on the left. The majority of stars in our data set, which follow disk-like orbits, form the highly populated right-hand envelope of the distribution. This confirms the picture from Figure~\ref{fig:toomre_rc_rgb} that most of our Li-rich stars are found in well-behaved disk orbits --- there are few Li-rich giants in orbits with high energy relative to their angular momentum, or with nonrotating or retrograde orbits. 

The dashed red rectangle in Figure~\ref{fig:lz_energy_rc_rgb} highlights the region of this parameter space occupied by the remnant of the \gaia-Enceladus merger event \citep{Helmi2018,Koppelman2018}. A handful of the Li-rich stars in our data set are located in this region of kinematic space. The recent study of \citet{Molaro2020} investigates the overall behaviour of lithium and beryllium in stars kinematically consistent with \gaia-Enceladus. Only one of the 101 stars in that study is Li-rich, which matches the 1\% occurrence rate we find in our overall sample. Simpson et al. 2021 (in prep.) expands on that work with a more detailed investigation to chemically and kinematically tag the stars in our data set to \gaia-Enceladus using data from GALAH DR3 and \gaia\ eDR3. The authors find that the overall evolution of lithium abundance is similar to the chemical evolution pattern in the Milky Way, suggesting that the Spite Plateau \citep{Spite1982} is a universal upper limit for lithium abundance in low-mass stars and is not a special feature of the Milky Way. Simpson et al. (2021) find four metal-poor halo stars in the halo of the Milky Way with lithium abundances at or above the primordial amount. 
 
\bsp	
\label{lastpage}
\end{document}